\documentclass[preprint]{aastex}

\begin{document}

\title{Redshifts for a Sample of Radio-Selected Poor Clusters}
\author{Neal A. Miller} 
\affil{NASA/Goddard Space Flight Center, LASP --- Code 681,\\
Greenbelt, MD  20771}
\email{nmiller@stis.gsfc.nasa.gov}

\author{Michael J. Ledlow}
\affil{Gemini Observatory, c/o AURA, Casilla 603,\\
La Serena, Chile}
\email{mledlow@gemini.edu}

\author{Frazer N. Owen}
\affil{National Radio Astronomy Observatory\altaffilmark{1}, P.O. Box O, \\ Socorro, New Mexico  87801}
\email{fowen@aoc.nrao.edu}

\author{John M. Hill}
\affil{Steward Observatory, University of Arizona\\
Tucson, AZ  85721-0065}
\email{jhill@as.arizona.edu}

\altaffiltext{1}{The National Radio Astronomy Observatory is a facility of the National Science Foundation operated under cooperative agreement by Associated Universities, Inc.}

\begin{abstract}
Multifiber optical spectroscopy has been performed on galaxies in the vicinity of strong, nearby radio galaxies. These radio galaxies were selected from the 3CR and B2 catalogs based on their exclusion from the Abell catalog, which is puzzling given the hypothesis that an external medium is required to confine the radio plasma of such galaxies. Velocities derived from the spectra were used to confirm the existence of groups and poor clusters in the fields of most of the radio galaxies. We find that all radio galaxies with classical FR I morphologies prove to reside in clusters, whereas the other radio galaxies often appear to be recent galaxy-galaxy mergers in regions of low galaxy density. These findings confirm the earlier result that the existence of extended X-ray emission combined with a statistical excess of neighboring galaxies can be used to identify poor clusters associated with radio galaxies.
\end{abstract}

\keywords{galaxies:active --- galaxies:clusters:general --- galaxies:distances and redshifts}

\section{Introduction}

Some of the more spectacular extragalactic sources are the powerful radio galaxies, with L$_{1.4GHz} \ge 10^{23}$ W Hz$^{-1}$. Associated almost exclusively with massive ellipticals, these objects often exhibit radio emission extending far beyond the optical extent of the host galaxies. The more extreme examples span over 1 Mpc with the radio emission resembling jets or lobes. In order to explain these large sizes, models require an external medium to confine the radio plasma and associated magnetic fields, and consequently the cluster environment with its intracluster medium (ICM) of hot gas is an attractive candidate. Indeed, many powerful radio galaxies are found to lie in clusters. However, not all powerful radio galaxies reside in known clusters of galaxies. How can such sources be explained?

One possible explanation is that such radio galaxies do, in fact, reside in clusters of galaxies but that these clusters are simply too poor to have been included in classical catalogs. For example, the lowest richness clusters in the Abell catalog \citep{abel1958,abel1989} must contain at least 35 galaxies in the interval described by the cluster's third brightest galaxy to a galaxy two magnitudes fainter than this galaxy. This arbitrary cutoff implies that systems which represent meaningful associations of galaxies could easily be missed, despite having environments nearly identical to those of the poorer clusters of the catalog.

Additionally, the connection between powerful radio galaxies and real galaxy overdensities might be exploited to identify samples of clusters and groups. The powerful radio galaxies would be nearly ideal signposts, as they are observable out to very high redshifts and are nearly always found in the centermost regions of clusters \citep{ledl1995}. In fact, studies including \citet{alli1993} have taken advantage of this technique. However, the picture is likely more complicated. Specifically, there is evidence that different types of powerful radio sources inhabit different environments. For example, radio galaxies are often separated based on radio luminosity and morphology \citep[and optical magnitude; see][]{owen1991,ledo1996} into FR I and FR II sources \citep{fana1974}. The FR I sources appear to be more frequently associated with cluster environments than FR II sources \citep[e.g.,][]{long1979,mill1999}, although this may only be true at lower redshift as at higher redshift their environments appear more similar \citep[e.g.,][]{pres1988}. There may also be environmental differences among the more powerful radio galaxies (generally FR II), with some being fairly normal optically luminous ellipticals whereas others are clearly disturbed and apparent galaxy-galaxy merger systems \citep{heck1986}. Consequently, in order to use radio galaxies to identify samples of clusters and groups it appears to be important to consider the nature of the radio source.

In \citet[][hereafter Paper 1]{mill1999} we explored these themes using a sample of nearby powerful radio galaxies which were not members of Abell clusters. The local environments of the radio galaxies were examined via analysis of X-ray data from the Rosat All-Sky Survey \citep[RASS; see][]{voge1993} and statistical measures of clustering from optical images. The X-ray data confirmed that the majority of the radio sources were associated with X-ray emission. We also noted a probable difference between FR I and FR II environments, with the FR I sources more frequently associated with extended X-ray emission and optical overdensities of galaxies. Thus, the FR I sources did appear to be associated with poor clusters while the FR II sources were more isolated.

In this paper, we report on the use of multifiber spectroscopy to further investigate the environments of powerful nearby radio galaxies which are not members of classical Abell clusters. The optical analysis employed in Paper 1 included no velocity information, and was instead based on the two-point spatial correlation function, $B_{gg}$. Velocity measurements are therefore necessary to confirm the presence of poor clusters around FR I radio galaxies, and further explore the environments of radio galaxies which appear to reside in very sparse regions of galaxies. The velocities are used to evaluate velocity dispersions in the identified poor clusters and estimate system masses under the assumption of virial equilibrium. In some cases, the obtained velocities also allow an investigation of potential substructure within the identified clusters and the relationship of the radio sources to such substructure. The goals are to: 1) confirm the existence of poor clusters identified through the presence of powerful radio galaxies, and 2) better understand the relationship between the characteristics of the radio galaxies and their local environments.

 The observations and reductions will be discussed briefly in Section \ref{sec:data}, followed by the computation of quantities to characterize their environments (Section \ref{sec:anal}). A discussion of the results is presented in Section \ref{sec:discuss}, followed by a brief summary of the conclusions. Throughout this paper, we have adopted $H_o=75$ km s$^{-1}$ Mpc$^{-1}$ and $q_o=0.1$ in all calculations which require such factors.

\section{Data}\label{sec:data}

\subsection{The Sample}

The complete sample is composed of 25 radio galaxies drawn from the 3CRR \citep{lain1983} and \citet{wall1985} catalogs, plus 24 additional radio galaxies drawn from the B2 catalog \citep{coll1970,coll1972}. The B2 sources are generally of lower radio luminosity, but were added as the declination range of that survey made them excellent targets for observation at Kitt Peak. The radio galaxies were required to be nearby ($z<0.06$ for the 3CRR and Wall \& Peacock galaxies, $z<0.04$ for the B2 galaxies), and not members of Abell clusters. We report on a subset of 25 of these radio galaxies in this paper. The remainder were unobservable due to declination limits of the telescope (6 sources) and observing time lost due to weather. The full sample of radio galaxies and comments on their properties may be found in Table \ref{tbl:sample}.

\placetable{tbl:sample}

\subsection{Observations and Reductions}

Spectra were obtained using the MX Spectrometer on the Steward Observatory 2.3 meter Bok telescope. MX utilizes 2\arcsec{} fibers on mechanical probes to obtain spectra of up to 32 targets at once, plus sky spectra from 30 fibers ``piggy-backed'' to target spectra probes. The field of view of the telescope with MX is 45\arcmin, which matches nicely with the angular sizes of the poor clusters being studied. In 18 of the 25 systems being studied this corresponds to a linear size of 1 Mpc or greater, while for the nearest system it is $\sim0.8$ Mpc. The actual spectrograph is the Steward Observatory Boller \& Chivens spectrograph, which rests on the dome floor. Using a 400 lines mm$^{-1}$ grating, 3300$\mbox{\AA}$ coverage is obtained at a resolution of about 6$\mbox{\AA}$. The wavelength center for the observations was typically around $5400\mbox{\AA}$, which covered from [OII]$\lambda3727$ to [SII]$\lambda\lambda6717+6731$ for the small redshifts of the sample. Additional details on the MX instrument may be found in \citet{hill1986,hill1988}.

Targets were selected from the Digitized Sky Survey plates, and corresponded to galaxies with magnitudes near that of the radio galaxy. In all but one case (B2 0120+33), the fields around the radio galaxies were sparse enough that all relatively bright galaxies could be observed through a single pointing of the telescope. The integration time for each pointing was one hour. Upon completion of each science exposure, comparison lamp arc spectra and continuum fiber flats (used to accurately locate the apertures corresponding to each fiber on the CCD) were obtained.

Data reduction was performed following the usual steps. On each night of observations, a sequence of 10 zero-second exposures and a sequence of 10 exposures of a quartz lamp (taken through a diffuser placed below the slit) were taken to create a master bias frame and flat field, respectively. The extracted spectra were wavelength calibrated using the arc lamp exposures, producing rms errors of under 0.2$\mbox{\AA}$ (determined from $\sim30$ lines and confirmed via inspection of airglow lines). The 30 sky fibers from each pointing were combined to create a super-sky spectrum (after rejecting any fibers which chanced to lie upon stars or galaxies), which was iteratively subtracted from each spectrum until the [OI] night sky line at $5577\mbox{\AA}$ was entirely removed.

\subsection{Inspection of Spectra and Velocity Measurements}

Each spectrum was carefully inpected in IRAF. Artifacts from cosmic rays and bad columns on the CCD were excised from the spectra, and velocity measurement then followed one of two procedures. First, some of the observed galaxies had spectra including emission lines such as the Balmer lines and forbidden oxygen, nitrogen, and sulfur lines. When such lines were present, velocities and errors were obtained using known laboratory wavelengths of such species. In these calculations, all lines were assigned equal weight in the determination of a galaxy's velocity and the estimate of the error in this quantity is simply the dispersion of the values obtained for individual lines. A minimum velocity error of 20 km s$^{-1}$ was adopted as an upper limit, based on the uncertainty of our wavelength solutions. For the majority of the galaxies, emission lines were not present or were too weak to be useful in velocity characterization. Radial velocities for such galaxies were determined via fourier cross-correlation \citep{tonr1979} using known velocity standards. A set of templates including M31, M32, NGC 3379, NGC 7331, and the brightest galaxies from several nearby clusters were used as the velocity standards, with the cross correlations using wavelengths between 3700$\mbox{\AA}$ and 6750$\mbox{\AA}$ (minus small regions centered on the prominent night sky lines at 5577$\mbox{\AA}$ and 6300$\mbox{\AA}$). Each galaxy spectrum was cross correlated against each template spectrum, and the resulting velocities were 3$\sigma$ clipped. Final velocities were assigned if at least $60\%$ of the templates produced a consistent velocity with an internal dispersion among the individual measurements that was less than 250 km s$^{-1}$. In general, the proximity of the targets resulted in $R\gtrsim8$ (see Tonry \& Davis). Velocity errors were assigned using the formula $280(1 + R)^{-1}$ \citep[Tonry \& Davis and][]{oege1998}, again with a minimum of 20 km s$^{-1}$. Finally, each galaxy velocity was confirmed via visual inspection through identification of major features such as the 4000$\mbox{\AA}$ break and prominent absorption features such as the G band, MgII, and NaD.

The relative proximity of the sample meant that velocities for some galaxies in the fields of the radio galaxies had already been measured. Consequently, we obtained velocities from the NASA/IPAC Extragalactic Database (NED) for galaxies within 1 Mpc of the radio galaxy positions. For galaxies in the field of 3C31 we have adopted the velocities of \citet{ledl1996}, since these observations were made using the same telescope and detector. 

Table \ref{tbl:memvel} presents the results for all galaxies deemed to be associated with the radio galaxies (see Section \ref{sec:anal}). The same information for foreground and background galaxies may be found in Table \ref{tbl:bkg_clus}. If emission lines were present in the spectrum, they are noted in the final column. When such lines were not used in the velocity calculation, they are offset by parentheses.

\placetable{tbl:memvel}

\placetable{tbl:bkg_clus}

\section{Analysis}\label{sec:anal}

\subsection{Velocity Completeness}

The primary goals of this study are to confirm the presence of poor clusters around the radio galaxies and better understand the connection of these galaxies to their local environments. For these purposes, the biases inherent in our sampling of galaxy velocities should not be too important. However, it is instructive to examine the completeness of our velocity data. Hence, we evaluate the fraction of galaxies with measured velocities as a function of optical magnitude and radial separation from the radio galaxies.

Galaxy identifications and magnitudes are available for 21 of the 25 radio galaxy fields in the Automated Plate Scanner (APS) catalogs \citep[for a description of the scanner and procedure, see][]{penn1993}. The other fields are generally too close to the Galactic plane, causing difficulty in star/galaxy segregation. We have obtained identifications and magnitudes for all galaxies located within a projected separation of 1 Mpc of the radio galaxies and used them to evaluate the completeness of our velocity data. Table \ref{tbl:complete} presents the results, including the limiting magnitude ($R_c$, derived from the Palomar ``E'' plates) below which we have velocity data for all galaxies (including those not associated with the radio galaxies). Information is provided for all galaxies within 1 Mpc of the radio galaxies, in addition to those within 500 kpc and 250 kpc. Our velocity information is more complete in the central regions of the radio galaxy fields, as these were more likely to be the target of our spectroscopic observations. For galaxies within 1 Mpc of the radio galaxy, our average limiting magnitude is 15.3 (for comparison, the radio galaxies range from $11.2\leq m_R \leq 14.6$). The fainter galaxies with spectra of high enough S/N to obtain accurate velocity measurements typically have $m_R\sim17$. Within 250 kpc of the radio galaxies, the limiting magnitude has dropped to an average of 16.1. Therefore, we conclude that the spectroscopy in these fields adequately samples the velocity fields in the vicinity of the radio galaxies.

\placetable{tbl:complete}

A related issue is the confirmation of the statistical results from Paper 1. In that study, counts of the brighter galaxies in the radio galaxy fields were used to evaluate $B_{gg}$. The velocity data confirm that for three of the six fields in common between these studies (3C 293, 3C 296, 1615+351; 3C 386 is too close to the Galactic plane to be included in the APS catalog) all of the galaxies used in the $B_{gg}$ calculations are cluster members. In a fourth (3C 31), nine out of ten are confirmed cluster members, a result consistent with the estimate of background galaxies in the $B_{gg}$ calculation. The remaining source (3C 305) had seven possible members in the $B_{gg}$ analysis, of which only three were cluster members. Thus, the statistical evidence for clustering in this radio galaxy field was misleading. However, 3C 305 was shown to not be associated with extended X-ray emission and was therefore a poor candidate for a true cluster system.

\subsection{Derived Quantities: Systemic Velocities, Velocity Dispersions, and Virial Masses}\label{sec:vels}

The systemic velocities and velocity dispersions were calculated using 3$\sigma$ rejection \citep{yahi1977}. Due to the low number of velocities available, we have adopted the Biweight estimators of location and scale ($C_{BI}$ and $S_{BI}$, respectively) instead of the simple mean and standard deviation \citep{beer1990}. These estimators are less sensitive to the presence of outliers, and are consequently better suited to the identification of real systems. The 1$\sigma$ confidence intervals in the biweight scale and location were estimated using the jackknife method \citep{most1977}. In all subsequent discussion, the terms systemic velocity and velocity dispersion will be used for $C_{BI}$ and $S_{BI}$. It should be noted that other algorithms exist; for example, the ``DEDICA'' method which uses the data to estimate the underlying probability density and thereby identify clusters and assign probabilities that individual galaxies are members \citep{pisa1993}.

The results may be found in Table \ref{tbl:sumtab}. Since the radio galaxies may exist in localized substructures within richer environments or the velocity dispersions might be artificially high should unassociated groups be evaluated together, all quantities have been determined for both a 1 Mpc radius and a 250 kpc radius centered on the radio galaxy. It should be noted that the values for the 250 kpc aperture have not had an additional 3$\sigma$ clipping performed, but were simply calculated using the culled lists derived from the 1 Mpc apertures. Figure \ref{fig-1} presents velocity histograms for all of the observed radio galaxy fields, using the 1 Mpc radius. In these figures, a range in velocity of $\pm$2000 km s$^{-1}$ with a bin size of 200 km s$^{-1}$ is shown. Verified cluster members are represented as the shaded portion of the histogram, with foreground and background galaxies unshaded. An overlaid Gaussian with center and dispersion corresponding to the systemic velocity and velocity dispersion of the system is also depicted for each field. Lastly, the arrows denote the locations of the radio galaxies.

\placefigure{fig-1}

\placetable{tbl:sumtab}

In two cases, the 3$\sigma$ clipping placed the radio galaxy outside of its apparent system. Most notably, B2 1322+36 was clipped from a system at 5,625 km s$^{-1}$. The overall system appears to consist of nine members, with six of the galaxies tightly grouped in velocity space. These six galaxies produce a velocity dispersion of only 30 km s$^{-1}$, placing the radio galaxy velocity of 5,210 km s$^{-1}$ outside of the identified group. Other methods would likely include all of these galaxies in the identified system; in fact, the system has also been noted in prior group studies \citep[e.g.,][]{whit1999} which included the radio galaxy among its members. Simple usage of the standard mean and deviation (including 3$\sigma$ clipping) produce a single nine galaxy system with $cz=5614$ km s$^{-1}$ and $\sigma = 200$ km s$^{-1}$. We therefore conclude that this is likely a loose system dominated by a group of galaxies strongly clustered in velocity space. The other radio galaxy which was clipped from a nearby system was 3C 305. It lies near a pair of galaxies at approximately the same velocity. 

The distributions and velocities of the galaxies were also used to estimate the masses of the potential clusters, under the assumption of virial equilibrium. The virial masses were calculated using the standard equation:
\begin{equation}
\mathcal{M}~=~\frac{3 \pi}{G} \sigma^2 \langle \frac{1}{r} \rangle ^{-1}
\end{equation}
where the harmonic mean radius is defined as
\begin{equation}
\langle \frac{1}{r} \rangle ^{-1} ~ = ~ D_L N_p \left( \sum_{i < j}^N \frac{1}{\theta_{ij}} \right)^{-1}
\end{equation}
and $N_p$ is the number of galaxy pairs ($N(N-1)/2$), $\theta_{ij}$ is the angular separation of each $ij$ pair, $D_L$ is the distance of the cluster, and $\sigma$ is the line of sight velocity dispersion. Errors in the velocity dispersions were propagated into the mass estimates. The resulting values were converted into units of solar masses and may be found in Table \ref{tbl:sumtab}. We have not applied any ``surface term correction'' to the mass estimates \citep{the1986} because this correction requires knowledge of the mass density profiles. The mass density profile would be obtained from some assumption of how well the galaxies trace the mass and how well the overall galaxy distribution is sampled. In addition, should substructure be present in the clusters (see Section \ref{sec:rgmove}) application of the virial theorem is obviously inaccurate. These factors tend to produce an overestimate of the true mass. Thus, the formal errors are larger than those quoted in Table \ref{tbl:sumtab}. Given these caveats, the masses are generally typical of groups and poor clusters, ranging up to $\sim3 \times 10^{14}$ M$_\odot$. 

\subsection{Simple Dynamical Analysis}\label{sec:rgmove}

A potentially important factor in assessing the environment of the radio galaxies is their locations within any surrounding clusters. Specifically, with the velocity data in hand we may ask whether the radio galaxies appear to be at rest within the clusters or have some relative motion. Relative motion between the host galaxy and the intracluster medium is required in order to explain certain radio morphologies, such as the bending of tails. In particular, narrow-angle tail (NAT) morphologies require substantial relative motion. However, in general we might expect the radio galaxies to be nearly at rest as they are often associated with the most massive elliptical galaxies in their fields. 

Using the systemic velocities and errors along with the velocities and errors for the radio galaxies, the significance at which the radio galaxy velocities differed from those of their host clusters was determined. Most of the radio galaxies were shown to lie at the velocity centers of their systems (15 of 25 within 1$\sigma$, including those systems with three or fewer velocities). Only three of the radio galaxies (B2 0222+36, B2 1321+31, and B2 1322+36) appear to have velocities which differ by more than 2$\sigma$ from the velocities of their assumed systems. The result is similar when assessing only those galaxies within 250 kpc of the radio sources (B2 0222+36, B2 1322+36, and 1615+351 had greater than 2$\sigma$ velocity offsets). It is perhaps not surprising that these three radio sources are the only ones in the sample which are not the brightest galaxies in their respective fields. However, the low number of velocities for most of the radio galaxy fields in the sample prevents any strong conclusions (see Table \ref{tbl:sumtab} for numbers of spectroscopically-identified velocities). 

Several of the identified groups have fairly large numbers of measured velocities. \citet{zab1998b} investigated six poor groups with large numbers of measured velocites ($N_{gal}\ge30$) to evaluate hierarchical evolution. Using the $\Delta$ test \citep{dres1988} they confirmed that evidence for substructure was present in two of the poor groups, arguing that these systems evolve hierarchically in much the same way as richer systems such as Abell clusters. Because the presence of substructure has been shown to be linked to radio activity in tailed radio galaxies \citep{gome1997,blit1998,pink2000}, we have applied substructure tests to those groups with more than 20 velocities: B2 0120+33 (65 velocities), B2 1621+38 (24 velocities), 3C31 (52 velocities), 3C296 (21 velocities), and 1615+351 (38 velocities).

A battery of statistical tests for substructure were performed \citep[for a complete listing of the tests performed see][PRBB96]{pink1996}. These tests include a number of normality tests \citep[i.e., comparison of the distribution of velocities with a normal distribution; see the ``ROSTAT'' routines of][]{beer1990}, tests for substructure in spatial distribution (right ascension and declination), and tests combining both velocity and spatial information. The significance of any detected substructure for the two- and three-dimensional tests was quantified by comparison against 1000 Monte Carlo shuffles of the actual data. Thus, each statistic calculated using the real data was compared to 1000 Monte Carlo simulations to evaluate how likely it occured relative to the null hypothesis of no substructure. For the two-dimensional tests, this null hypothesis is a smooth, azimuthally-symmetric distribution of galaxies. No correlation between galaxy position and velocity is the null hypothesis for the three-dimensional tests. The quoted significance level represents how often the test result for the actual data had more substructure than the Monte Carlo shuffles. We adopted a level of $99\%$ as significant (as per the conclusions of PRBB96), or fewer than 1 out of every 100 Monte Carlo simulations showing greater substructure than the real data.

For the most part, evidence for significant substructure was not found. For B2 0120+33, the 2D and 3D Lee tests \citep[][PRBB96]{lee1979,fitc1988} were each significant at about $95\%$, and the Angular Separation Test \citep{west1988} was significant at about $98\%$. In total, this represents marginal evidence for the presence of substructure in this cluster. Similarly, 3C 31 had a $\beta$ test \citep{west1988} that was significant at $98\%$ and B2 1621+38 had an Angular Separation Test signficant at $99\%$. No other tests were above $90\%$ significance, as was the case for all tests applied to 3C 296. The strongest evidence for substructure was found in 1615+351. Its galaxy distribution appears to be elongated, as the Fourier Elongation Test (PRBB96) was significant at well over $99\%$ confidence. Furthermore, its Lee 2D and 3D tests were also significant at about the $99\%$ level. While the Fourier test merely identifies an elongated distribution of galaxies which may or may not be caused by substructure and merging, the Lee tests are insensitive to such distributions. Hence, the combination of these tests argue for the presence of real substructure. The normality tests (based only on velocities) did not show evidence for substructure, although the velocity of the radio galaxy differed from that of the parent cluster by 1.9$\sigma$ (3.4$\sigma$ from the galaxies within 250 kpc).

\section{Discussion}\label{sec:discuss}

The majority of the radio galaxies appear to reside in poor clusters or groups. Nineteen of 25 examined fields consisted of at least five galaxies with velocities placing them in a system including the radio galaxy (twenty if we include B2 1322+36). In fact, over half of the fields (14 of 25) had in excess of 10 members, and five had more than 20 members. Two of the radio galaxies, B2 0120+33 and 3C 31, were shown to reside in very rich systems with 65 and 52 confirmed velocities, respectively. 

\begin{figure}
\figurenum{2}
\epsscale{0.85}
\plotone{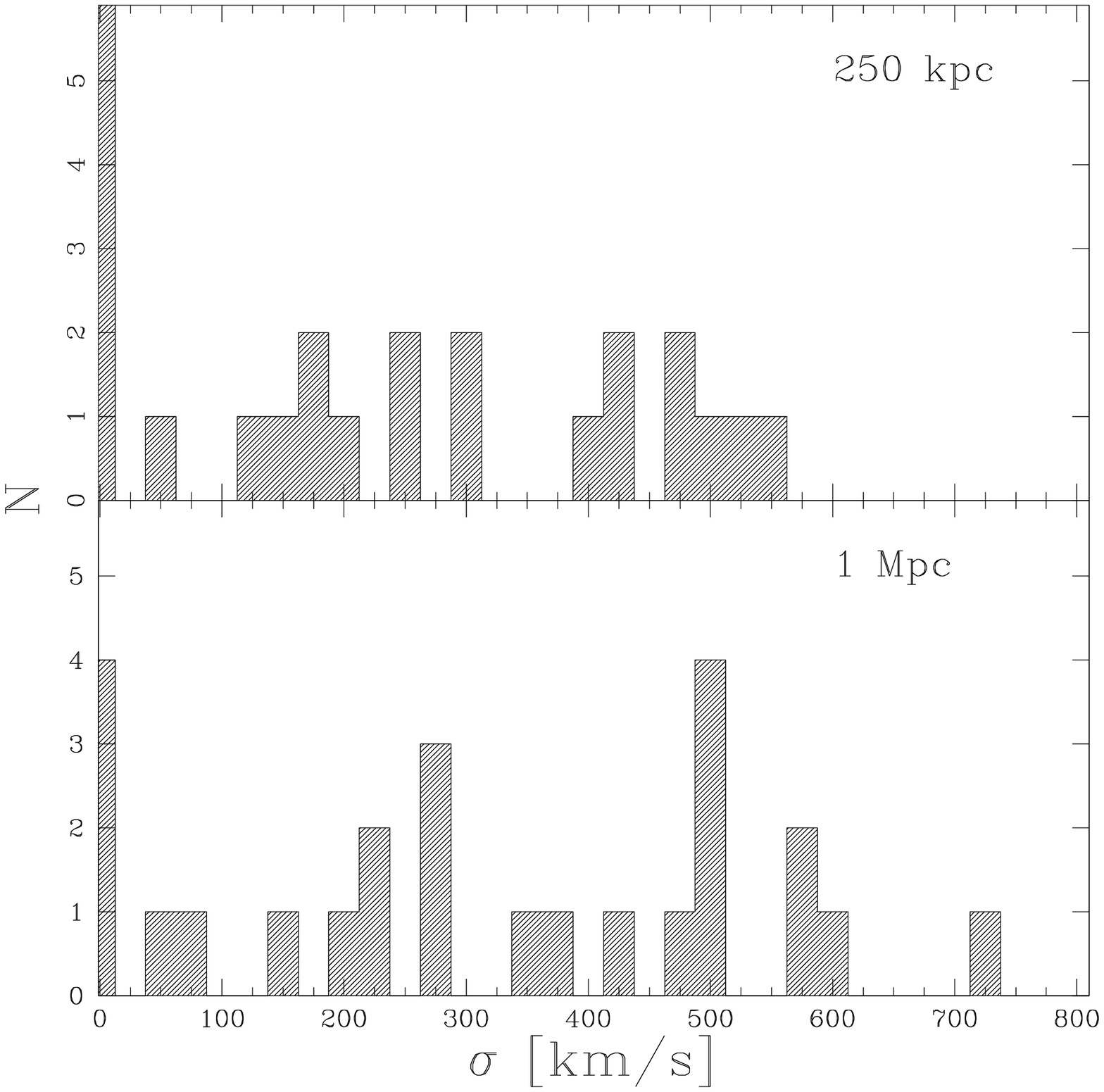}
\caption{Distribution of the derived velocity dispersions for the sample. The top portion represents the dispersions calculated with 250 kpc (projected) of the radio galaxy, while the bottom represents those calculated for galaxies within 1 Mpc. \label{fig-2}}
\end{figure}

The calculated velocity dispersions and virial masses further indicate that the radio galaxies tend to exist in groups and poor clusters. Figure \ref{fig-2} depicts histograms of the derived velocity dispersions for each the 250 kpc and 1 Mpc counting radii. About a third of the fields have dispersions around 200 km s$^{-1}$, consistent with the values found for nearby groups \citep[e.g., nearly 400 loose groups identified from the Las Campanas Redshift Survey have a median velocity dispersion of 164 km s$^{-1}$ and a median mass of about $2.5 \times 10^{13}$ M$_\odot$;][]{tuck2000}. Most of the radio-selected groups in the present study have velocity dispersions and masses in excess of these values, and more consistent with the velocity dispersions found for groups and poor clusters \citep[e.g.,][]{ledl1996,zabl1998}. The overall distribution of velocity dispersions is similar to that found by Ledlow et al., who found a median dispersion of $295 \pm 31$ km s$^{-1}$ for a set of optically-selected poor clusters. In the present sample, we find a median of $452 \pm 161$ km s$^{-1}$ (using only the 20 clusters with five or more velocities, to be consistent with Ledlow et al.). While these values indicate the presence of slightly richer environments around the radio galaxies, it must be noted that Ledlow et al. used a smaller 0.5 Mpc counting radius. The richer among the radio-selected groups have dispersions and masses in line with the poorer Abell clusters. The median velocity dispersion of the richness class 0 clusters reported in \citet{zabl1991} is around 500 km s$^{-1}$, a value which is exceeded by seven of the radio groups in our sample. The median velocity dispersion for richness class 1 clusters in the same study is about 725 km s$^{-1}$, which is in excess of any of the velocity dispersions for our radio galaxy fields. Thus, the radio galaxies generally reside in structures ranging from groups on up through the poorer Abell clusters. 

\placefigure{fig-2}

Galaxy groups have received a great deal of attention in the literature as their high local densities and low velocity dispersions make them attractive sites for galaxy evolution. It is therefore not surprising that many of the radio galaxy fields have been associated with groups in prior studies. For example, eight of the 25 fields studied in this work are noted in \citet[][``WBL'' clusters]{whit1999}, and a similar number are reported in \citet[][the ``RASSCALS'']{mahd2000}. It is also interesting to note that the radio galaxy poor clusters are frequently associated with larger-scale structures. At least four are associated with the Pisces-Perseus supercluster and another two are associated with known groups at the periphery of Abell 2197 and 2199. Even with these prior identifications, a number of the radio-selected poor clusters appear to be new. Our data more than double the number of velocities for six of the fields which previously had five or fewer publicly-available velocities. The most striking example is 3C 296, to which we have added sixteen cluster velocities to the previously known five.

\begin{figure}
\figurenum{3}
\epsscale{0.85}
\plotone{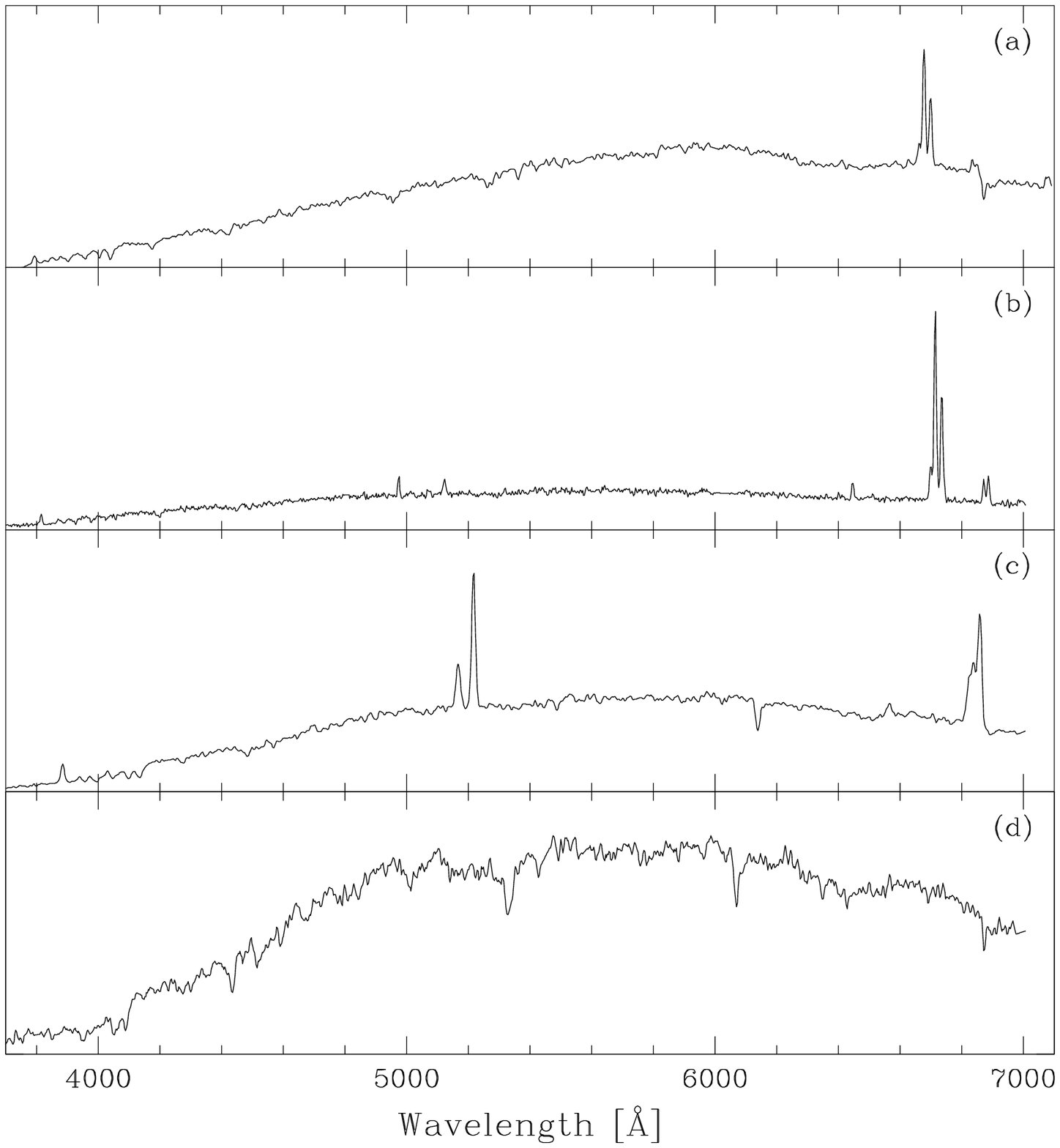}
\caption{Representative spectra of radio galaxies: (a) B2 0207+38, a dusty starburst; (b) B2 1318+34, a starburst; (c) 3C 305, a strong emission-line AGN; and (d) 1615+351, a radio galaxy dominated by an old stellar population. The majority of the observed radio galaxies have spectra resembling (d). \label{fig-3}}
\end{figure}

However, five of the fields produced three or fewer member velocities, indicating that these radio galaxies are almost certainly not in groups or poor clusters. How can these objects be explained? As noted previously, the nature of the radio source is an additional parameter of importance (refer to Table \ref{tbl:sample}). Two of the five (B2 0207+38 and B2 1318+34) are powered by star formation and not an AGN. Optical images show them to be spirals, and their spectra (see Figure \ref{fig-3}a and \ref{fig-3}b) show strong emission of H$\alpha$. In fact, \citet{pogg2000} discuss B2 0207+38 as a nearby example of a strong dust-extincted starburst and \citet{cond1991} demonstrate that B2 1318+34 is a merger-induced starburst. The remaining three radio galaxies in very poor environments are all AGN, but differ from the more typical radio sources of the sample. Both 3C 293 and 3C 305 are galaxy-galaxy merger systems \citep{heck1986,evan1999}, with strong dust features and signs of disturbance \citep{mart1999}. Most of the radio emission from 3C 293 is contained within a bright central core and all of the radio emission from 3C 305 is contained within the optical extent of the galaxy, making the radio morphologies of these sources distinctly different from classical FR I galaxies. They also exhibit evidence for star formation, including large amounts of far-infrared emission, the presence of large amounts of molecular gas \citep{evan1999}, and emission lines in their optical spectra (see Figure \ref{fig-3}c). The remaining radio galaxy which appears to be fairly isolated is 3C 386. Although the arguments are weaker, it has some similarities with 3C 293 and 3C 305. Its radio morphology is that of a relaxed double, with a bright core surrounded by diffuse emission over a large area. Consequently, it is not a classical FR I type radio source.

\placefigure{fig-3}

An additional three of the radio galaxies have LINER spectra (B2 0222+36, B2 0258+35, and B2 1422+26). In general, these are also located in poorer environments than the classical FR I radio sources. Both B2 0258+35 and B2 1422+26 have only five confirmed members, and B2 0222+36 resides in a system with 14 confirmed members. However, the local environment of this latter source may be more complicated. It is not the brightest galaxy in its field, and appears to have a velocity which is offset from systemic velocity. The small number of velocities prohibits any statistical statements, but this system might be a pair of groups --- one which includes the radio galaxy and another which includes the brighter galaxy.

The rest of the radio galaxies generally show classical FR I morphologies, including jets of radio emission which extend past the optical boundaries of the galaxies. Optically, they all have absorption-line spectra indicative of older stellar populations. With the exception of B2 1322+36, all of these galaxies have velocity dispersions in excess of 200 km s$^{-1}$ indicating their presence in structures ranging from the size of groups on up through poor Abell clusters, with virial masses from about $2 \times 10^{13}$ to $3 \times 10^{14}$ M$_\odot$. This result is in agreement with the conclusions of Paper 1: the presence of an FR I radio source is an excellent indicator of an underlying group or poor cluster\footnote{As an aside we note that B2 1652+39, often classified as a BL Lac, resides in an environment consistent with the classical FR I sources of the sample. This is the expected result from FR I--BL Lac unification scenarios.}. 

\begin{figure}
\figurenum{4}
\epsscale{0.85}
\plotone{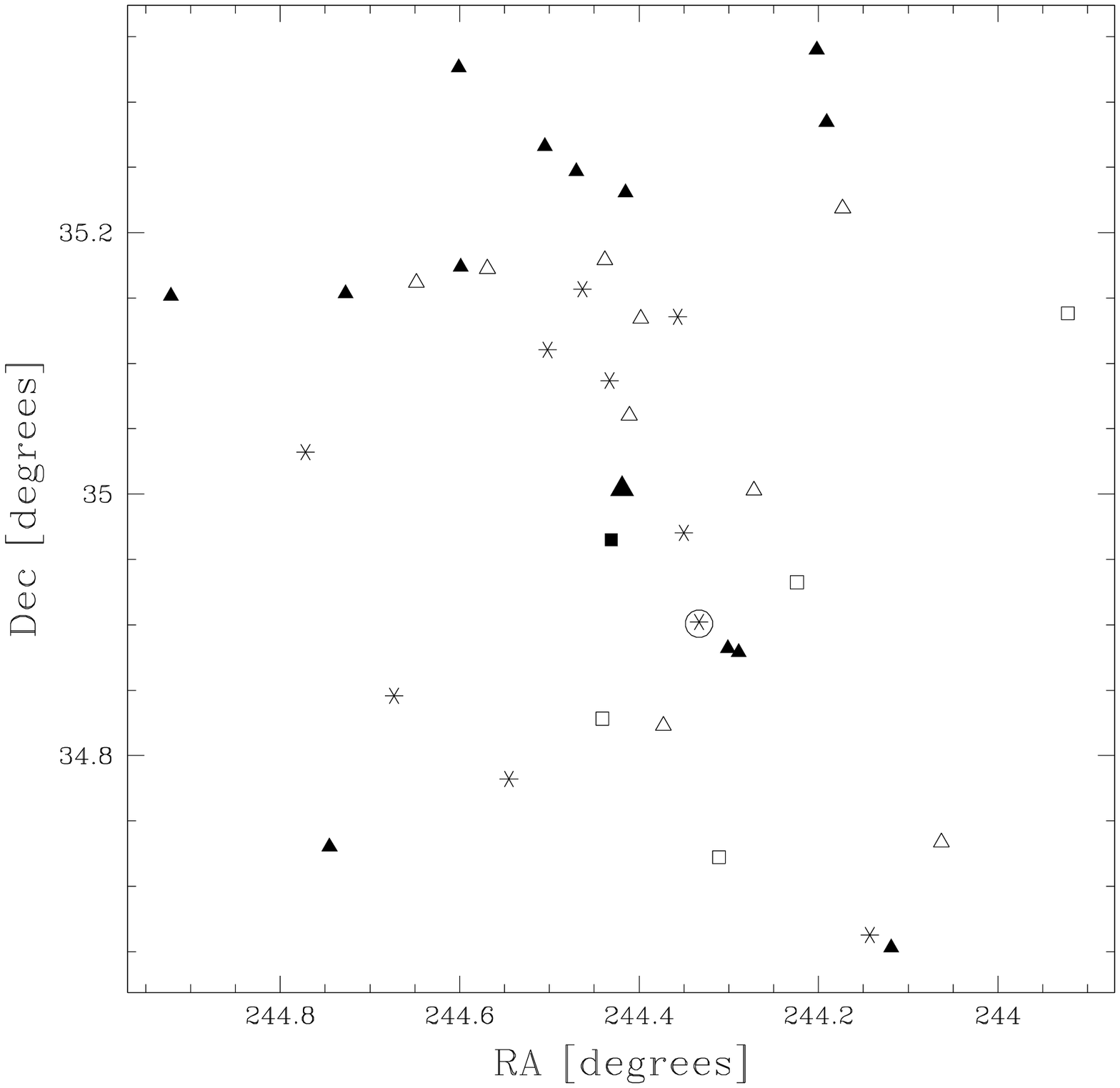}
\caption{Galaxy distribution for 1615+351. The symbols represent velocity ranges for the galaxies: filled squares, $cz < 8410$ (km s$^{-1}$); filled triangles, $8410 \leq cz < 8994$; asterisks, $8994 \leq cz \leq 9578$; open triangles, $9578 < cz \leq 10162$; and open squares, $cz > 10162$. The radio galaxy (NGC 6109) is at the center of the field and is denoted by a larger symbol. Its velocity is 8979 km s$^{-1}$. The brightest galaxy in the field (NGC 6107) has been circled; it is 0.3 magnitudes brighter than the radio galaxy and has a velocity of 9399 km s$^{-1}$. \label{fig-4}}
\end{figure}

Two of the FR I radio sources (B2 1621+38 and 1615+351) have NAT morphologies. They are located in two of the richer environments of the sample, and each have associated diffuse X-ray emission \citep{fere1995}. \citet{blit1998} examined NAT sources drawn from a large sample of Abell clusters and noted that clusters with NATs were more likely to have substructure than radio-quiet clusters. Coalescence of substructures can produce the large relative motions between NAT host galaxies and the intracluster gas necessary to explain the radio morphologies of such sources. For B2 1621+38, there is only slight evidence for substructure, a result undoubtedly related to the somewhat limited number of velocites (24) in our study. The situation is improved for 1615+351, with its 38 velocities producing several significant results (see Section \ref{sec:rgmove}). In addition, the radio galaxy is the second brightest galaxy in the field, with the brightest being NGC 6107. This galaxy is also a radio source \citep{eker1978}, although it is weaker than 1615+351. Figure \ref{fig-4} plots the galaxy distribution for 1615+351. The elongation is evident, running from SW to NE \citep[which is also consistent with the diffuse X-ray emission;][]{fere1995}. A clump of galaxies at lower velocities than the cluster center is seen to the NE. For reference, the radio emission points away from the galaxy toward the NW. In total, these results indicate that the NATs in our study appear to be found in clusters with substructure, consistent with the findings for NATs in Abell clusters.

\placefigure{fig-4}

In general, the stronger radio sources (i.e., after removing star forming galaxies) can be split into two idealized categories as initially suggested by \citet{heck1986}. The first class includes the more powerful radio galaxies, with radio emission concentrated in a central core but often including diffuse emission on larger scales. These galaxies frequently have optical emission lines and reside in regions of low galaxy density. They often exhibit signs of disturbance related to recent galaxy-galaxy merging, presumably with at least one of the merger partners a disk galaxy which provides a reservoir of gas to fuel the AGN. The second class are FR I type sources, which appear more like typical large ellipticals and reside in regions of high galaxy density. Their radio emission extends well past the optical limits of the host galaxies in distinct jet-like structures. Of course, the situation is more complex as not all radio galaxies can easily be pigeon-holed into these two categories.

Unfortunately, our present sample includes no clear FR II galaxies with bright lobes of radio emission. The infrequency of such sources implies that there are very few within the redshift cutoffs we have applied. Such galaxies appear to reside in regions of lower galaxy density, at least for the low redshifts examined in this study. In the future, a comprehensive study of the environments of nearby FR II sources would do much to help understand the possible differences in the environments of FR I and FR II radio galaxies.

The placement of fibers on any galaxies within the fields of the radio galaxies also produced a number of velocities for apparent background groups and clusters. The presence of such systems may be inferred from Table \ref{tbl:bkg_clus}. As an example, the observation of B2 1652+39 appears to have netted seven velocities for the cluster Abell 2235. The cluster is presumably located at $\alpha = $16:54:58 and $\delta = $40:01:16 \citep{abel1989} which is under 22 minutes from the field center of our observations. \citet{stru1999} report that this cluster has a systemic velocity of 45,300 km s$^{-1}$, determined from three published velocities. From our seven velocities, we find a systemic velocity of 44,454$\pm$120 km s$^{-1}$ with a dispersion of 197$^{+149}_{-85}$ km s$^{-1}$. There is also extended X-ray emission associated with this cluster, as indicated by its presence in cluster list generated from the ROSAT All-Sky Survey \citep[identified as RX J1652.6+4011;][]{bohr2000}.

\section{Conclusions}

In this paper, we have confirmed that powerful radio galaxies are excellent signposts to the presence of clusters of galaxies. Using a sample of 25 radio galaxies drawn from the 3CRR and B2 catalogs, our velocity measurements determined that $80\%$ (20/25) consisted of at least five galaxies. Over half of the fields (14/25) had in excess of 10 members, with some having as many as 50 members. The derived velocity dispersions are consistent with those derived for groups and poor clusters. In particular, we note that radio galaxies with classical FR I morphologies seem to reside exclusively in group and poor cluster environments. The few radio galaxies which exist in regions of particularly low galaxy density are either vigorous star-forming galaxies or complex galaxy-galaxy mergers. These results confirm our prior findings which were based on statistical measures of galaxy clustering and the presence of extended X-ray emission.

The small sample of radio galaxy fields examined here provides additional clues regarding the formation and evolution of powerful radio galaxies. When the radio galaxies powered by star formation are removed from the sample, there are two general types of radio sources. The first class, which consists of more classical FR I radio galaxies which exhibit clear jets of radio emission, are found in regions of increased galaxy density. These regions have intracluster gas, as inferred through X-ray observations, and this gas can confine the radio plasma emitted by the radio galaxy. The second class of radio galaxy resides in regions of low galaxy density. They often exhibit optical emission lines and evidence for recent merger activity. Thus, environment seems to play a key role in determining the class of the radio galaxy. This picture is clearly an oversimplification, but generally explains our results. One important further area of study is how more classical FR II sources, with large lobes of bright radio emission, fit into this picture.

\clearpage

\acknowledgements
The authors thank Dennis Means and Vic Hansen for their assistance during the observations, Jason Pinkney for providing the computer code used in the analysis of possible substructures, and an anonymous referee for comments which substantially improved the focus of the paper. NAM thanks the National Radio Astronomy Observatory for financial assistance in the form of a predoctoral scholarship.

The Digitized Sky Survey was produced at the Space Telescope Science Institute under U.S. Government grant NAG W-2166. The images of these surveys are based on photographic data obtained using the Oschin Schmidt Telescope on Palomar Mountain and the UK Schmidt Telescope. The plates were processed into the present compressed digital form with the permission of these institutions. The Oschin Schmidt Telescope is operated by the California Institute of Technology and Palomar Observatory. The UK Schmidt Telescope was operated by the Royal Observatory Edinburgh, with funding from the UK Science and Engineering Research Council (later the UK Particle Physics and Astronomy Research Council), until 1988 June, and thereafter by the Anglo-Australian Observatory. The blue plates of the southern Sky Atlas and its Equatorial Extension (together known as the SERC-J), as well as the Equatorial Red (ER), and the Second Epoch [red] Survey (SES) were all taken with the UK Schmidt.

This research has made use of the APS Catalog of POSS I, which is supported by the National Aeronautics and Space Administration and and the University of Minnesota. The APS databases can be accessed at http://aps.umn.edu/. NED is operated by the Jet Propulsion Laboratory, California Institute of Technology, under contract with the National Aeronautics and Space Administration.

\begin{deluxetable}{l c c c c l c}
\tablecolumns{7}
\tablecaption{3C and B2 Non-Abell Cluster Radio Galaxies \label{tbl:sample}}
\tabletypesize{\tiny}
\tablewidth{490pt}
\tablehead{
\colhead{Source} & \colhead{RA(J2000)} & \colhead{Dec(J2000)} & 
\colhead{$z$} & \colhead{Observed?} & \colhead{Comments} & \colhead{Ref}
}
\startdata
B2 0034+25   & 00:37:05 & +25:41:56 & 0.032 & Y & FR I, Wide-angle tail & 2 \\
B2 0055+30   & 00:57:49 & +30:21:09 & 0.016 & Y & FR I, core and jets & 3 \\
3C 31        & 01:07:25 & +32:24:45 & 0.017 & Y & FR I & 1 \\
B2 0120+33   & 01:23:40 & +33:15:20 & 0.016 & Y & FR I, steep spectrum; faded? & 4,5 \\
PKS 0131-36  & 01:33:58 & -36:29:36 & 0.030 & N\tablenotemark{a} & FR II & 1 \\
B2 0206+35   & 02:09:39 & +35:47:50 & 0.037 & Y & FR I, core and jets & 5 \\
B2 0207+38   & 02:10:10 & +39:11:25 & 0.018 & Y & Star formation, spiral & 5 \\
3C 66B       & 02:23:11 & +42:59:32 & 0.022 & N & FR I & 1 \\
B2 0222+36   & 02:25:27 & +37:10:28 & 0.033 & Y & Flat spectrum & 5 \\
B2 0258+35   & 03:01:42 & +35:12:21 & 0.016 & Y & Flat spectrum & 5 \\
3C 76.1      & 03:03:15 & +16:26:19 & 0.032 & N & FR I & 1 \\
3C 78        & 03:08:26 & +04:06:39 & 0.029 & N & FR I & 1 \\
3C 88        & 03:27:55 & +02:33:44 & 0.030 & N & FR II & 1 \\
B2 0326+39   & 03:29:25 & +39:47:54 & 0.024 & Y & FR I & 6 \\
B2 0331+39   & 03:34:18 & +39:21:25 & 0.020 & Y & Core plus faint halo & 5 \\
3C 98        & 03:58:55 & +10:25:47 & 0.031 & N & FR II & 1 \\
PKS 0453-20  & 04:55:24 & -20:34:13 & 0.035 & N & FR I & 1 \\
Pictor A     & 05:19:49 & -45:46:46 & 0.035 & N\tablenotemark{a} & FR II & 1 \\
B2 0722+30   & 07:25:37 & +29:57:15 & 0.019 & N & Star formation, spiral & 5 \\
DA 240       & 07:48:37 & +55:48:59 & 0.036 & N & FR I/II & 1 \\
B2 0924+30   & 09:27:24 & +29:55:30 & 0.025 & N & Relic? & 13 \\
B2 1040+31   & 10:43:18 & +31:31:02 & 0.036 & N & Peculiar & 5 \\
B2 1108+27   & 11:11:25 & +26:57:46 & 0.033 & N & FR I & 5 \\
B2 1122+39   & 11:24:44 & +38:45:46 & 0.007 & N & FR I & 5 \\
3C 278       & 12:54:37 & -12:33:23 & 0.015 & N & FR I & 1 \\
B2 1317+33   & 13:20:18 & +33:08:41 & 0.038 & Y & Diffuse & 5 \\
B2 1318+34   & 13:20:35 & +34:08:22 & 0.023 & Y & Starburst & 7 \\
PKS 1318-434 & 13:21:09 & -43:42:39 & 0.011 & N\tablenotemark{a} & FR I & 1 \\
B2 1321+31   & 13:23:45 & +31:33:56 & 0.016 & Y & FR I, jets & 6 \\
B2 1322+36   & 13:24:51 & +36:22:42 & 0.018 & Y & FR I & 5 \\
3C 293       & 13:49:41 & +31:02:31 & 0.045 & Y & FR I/II & 12 \\
3C 296       & 14:16:53 & +10:48:27 & 0.024 & Y & FR I & 1 \\
B2 1422+26   & 14:24:41 & +26:37:30 & 0.037 & Y & FR I & 11 \\
3C 305       & 14:49:21 & +63:16:14 & 0.042 & Y & FR I/II, radio confined within galaxy & 12 \\
B2 1447+27   & 14:49:28 & +27:46:50 & 0.031 & Y & FR I & 11 \\
3C 310       & 15:04:57 & +26:00:59 & 0.054 & N & FR I/II & 1 \\
1615+351     & 16:17:41 & +35:00:16 & 0.030 & Y & FR I, Narrow-angle tail & 1,9 \\
B2 1621+38   & 16:23:03 & +37:55:20 & 0.031 & Y & FR I, Narrow-angle tail & 8,9 \\
NGC 6251     & 16:32:32 & +82:32:16 & 0.024 & N\tablenotemark{b} & FR I/II & 1 \\
PKS 1637-77  & 16:44:16 & -77:15:35 & 0.043 & N\tablenotemark{a} & FR II & 1 \\
B2 1652+39   & 16:53:52 & +39:45:37 & 0.034 & Y & Flat spectrum, BL Lac & 5,10 \\
3C 353       & 17:20:32 & -00:58:46 & 0.030 & N & FR II & 1 \\
3C 386       & 18:38:26 & +17:11:49 & 0.018 & Y & FR I, relaxed double & 1 \\
B2 2116+26   & 21:18:33 & +26:26:49 & 0.016 & Y & FR I, jets plus diffuse & 5 \\
PKS 2153-69  & 21:57:07 & -69:41:33 & 0.027 & N\tablenotemark{a} & FR I & 1 \\
3C 442A      & 22:14:47 & +13:50:27 & 0.017 & N & FR I & 1 \\
3C 449       & 22:31:21 & +39:21:30 & 0.017 & N & FR I & 1 \\
B2 2236+35   & 22:38:29 & +35:19:47 & 0.028 & Y & FR I, jets & 4 \\
PKS 2247+11  & 22:49:55 & +11:36:33 & 0.024 & N & FR I & 1 \\
\enddata

\tablenotetext{a}{Below declination limit of telescope.}

\tablenotetext{b}{Above declination limit of telescope.}

\tablecomments{Columns: (1) Radio galaxy; (2) Right Ascension (J2000); (3) Declination (J2000); (4) Redshift; (5) Flag indicating whether spectroscopic observations of the radio galaxy field are included in this paper; (6) Radio morphology and brief comments, (7) Reference for the radio morphology: 1 -- \citet{owen1989}, 2 -- \citet{deru1986}, 3 -- \citet{vent1993}, 4 -- \citet{cano1999}, 5 -- \citet{parm1986}, 6 -- \citet{liu1992}, 7 -- \citet{cond1991}, 8 -- \citet{odea1985}, 9 -- \citet{fere1995}, 10 -- \citet{laur1999}, 11 -- \citet{cape2000}, 12 -- \citet{mart1999}, 13 -- \citet{giov1988}.}
\end{deluxetable}

\begin{deluxetable}{c c c c c c c c}
\tablecolumns{8}
\tablecaption{3C/B2 Cluster Member Velocities \label{tbl:memvel}}
\tabletypesize{\small}
\tablewidth{475pt}
\tablehead{
\colhead{} & \colhead{} & \multicolumn{2}{c}{This Study} &
\colhead{} & \multicolumn{2}{c}{Prior Studies} & \colhead{} \\
\cline{3-4} \cline{6-7} \\
\colhead{Field} & \colhead{Position} & \colhead{$cz$} &
\colhead{Error} & \colhead{} & \colhead{$cz$} & \colhead{Error} & 
\colhead{Emission Lines}
}
\startdata
\sidehead{B2 0034+25}
 & 00:35:39.1 25:40:44 & \phn9838 & \phn22 & & & & \\
 & 00:35:58.3 25:40:54 & 10272 & \phn29 & & & & (H$\alpha$, [NII]) \\
 & 00:36:05.0 25:49:44 & 10420 & \phn22 & & & & \\
 & 00:36:08.7 25:45:04 & \phn9944 & \phn20 & & & & \\
 & 00:36:19.3 25:48:27 & \phn9664 & \phn20 & & 9760 & \phn35 & ([NII]) \\
 & 00:36:32.4 25:45:19 & 10570 & \phn20 & & & & \\
 & 00:36:56.5 25:42:57 & \phn9238 & \phn20 & & & & \\
 & {\it 00:37:05.5 25:41:56} & & & & \phn9548 & \phn15 & \\
 & 00:37:08.3 25:41:35 & & & & \phn9667 & \phn47 & \\
 & 00:37:11.7 25:50:22 & & & & \phn9014 & \phn\phn9 & \\
 & 00:37:14.6 25:43:11 & & & & 10456 & \phn31 & \\
\sidehead{B2 0055+30}
 & 00:54:59.0 30:48:20 & & & & \phn4684 & \phn\phn7 & \\
 & 00:55:22.7 30:29:13 & & & & \phn5149 & \phn10 & \\
 & 00:56:09.2 31:04:29 & & & & \phn4666 & \phn\phn7 & \\
 & 00:56:34.3 30:53:31 & & & & \phn4737 & 250 & \\
 & 00:57:32.3 30:16:51 & \phn4999 & \phn44 & & \phn5065 & \phn14 & \\
 & 00:57:32.1 30:30:17 & \phn5666 & \phn76 & & & & \\
 & {\it 00:57:48.8 30:21:30} & \phn4972 & \phn33 & & \phn4956 & \phn17 & \\ 
 & 00:58:01.2 30:42:18 & & & & \phn4723 & \phn\phn9 & \\
 & 00:58:05.2 30:25:32 & & & & \phn5307 & \phn28 & \\
 & 01:00:32.5 30:47:50 & & & & \phn4825 & \phn\phn9 & \\
 & 01:00:36.5 30:40:07 & & & & \phn4778 & \phn\phn6 & \\
 & 01:00:38.5 30:18:59 & & & & \phn5186 &           & \\
 & 01:01:01.2 29:56:51 & & & & \phn4884 & \phn\phn5 & \\
\sidehead{B2 0120+33}
 & 01:19:54.3 32:56:55 & & & & \phn4408 & \phn60 & \\
 & 01:20:01.7 33:39:21 & & & & \phn3844 & \phn78 & \\
 & 01:20:06.9 33:11:06 & & & & \phn5717 & \phn10 & \\
 & 01:20:13.1 33:30:23 & & & & \phn5190 & \phn36 & \\
 & 01:20:28.7 32:42:36 & & & & \phn5312 & \phn22 & \\
 & 01:20:37.8 33:33:26 & & & & \phn5334 & \phn78 & \\
 & 01:20:46.3 33:02:42 & & & & \phn5090 & \phn\phn7 & \\
 & 01:20:53.1 33:25:32 & & & & \phn5468 & \phn16 & \\
 & 01:21:03.0 33:53:55 & & & & \phn4137 & \phn33 & \\
 & 01:21:05.6 33:22:44 & & & & \phn5252 & \phn18 & \\
 & 01:21:07.1 33:12:58 & & & & \phn5417 & \phn\phn9 & \\
 & 01:21:17.4 33:05:26 & & & & \phn5141 & \phn\phn8 & \\
 & 01:21:18.3 33:09:31 & & & & \phn4086 & \phn62 & \\
 & 01:21:34.9 33:35:59 & \phn4471 & \phn29 & & \phn4489 & \phn34 & (H$\alpha$, H$\beta$, [NII]) \\ 
 & 01:21:37.4 32:36:21 & & & & \phn4745 & \phn39 & \\
 & 01:21:44.6 33:29:35 & \phn5421 & \phn47 & & \phn5310 & \phn28 & \\ 
 & 01:21:51.2 33:16:56 & \phn4469 & \phn20 & & \phn4438 & \phn\phn6 & \\
 & 01:21:55.6 33:07:51 & & & & \phn5366 & \phn77 & \\
 & 01:21:56.4 33:31:16 & \phn4666 & \phn20 & & \phn4678 & \phn22 & \\ 
 & 01:22:00.9 33:30:35 & & & & \phn5184 & \phn27 & \\
 & 01:22:11.8 33:26:55 & \phn4293 & \phn40 & & \phn4354 & \phn27 & \\
 & 01:22:13.1 33:15:36 & \phn4330 & \phn20 & & \phn4391 & \phn25 & \\
 & 01:22:23.5 33:48:55 & & & & \phn5129 & \phn37 & \\
 & 01:22:25.7 32:41:51 & & & & \phn5333 & \phn26 & \\
 & 01:22:29.2 34:02:24 & & & & \phn4916 & \phn79 & \\
 & 01:22:34.0 33:06:12 & & & & \phn6039 & \phn60 & \\
 & 01:22:48.7 33:58:06 & & & & \phn5505 & \phn30 & \\
 & 01:22:51.8 33:31:00 & \phn4290 & \phn20 & & \phn4132 & 100 & \\
 & 01:22:53.2 33:24:47 & \phn4018 & \phn20 & & \phn3952 & \phn58 & \\
 & 01:22:55.4 33:10:26 & \phn5469 & \phn20 & & \phn5462 & \phn\phn9 & (H$\alpha$, [NII], [OII], [OIII]) \\
 & 01:22:56.4 33:28:09 & \phn4124 & \phn42 & & \phn4114 & \phn10 & \\
 & 01:23:01.8 34:08:03 & & & & \phn5125 & \phn\phn7 & \\
 & 01:23:06.6 33:11:22 & & & & \phn4462 & \phn37 & \\
 & 01:23:11.3 33:29:22 & & & & \phn6151 & 100 & \\
 & 01:23:11.6 33:27:39 & \phn4365 & \phn20 & & \phn4399 & \phn10 & \\
 & 01:23:11.6 33:31:43 & \phn6151 & \phn51 & & \phn6006 & \phn\phn8 & (H$\alpha$, H$\beta$, [NII], [SII]) \\
 & 01:23:14.8 33:33:44 & \phn4498 & \phn53 & & \phn4424 & \phn15 & \\
 & 01:23:19.0 33:16:40 & \phn5019 & \phn20 & & \phn4881 & \phn60 & \\
 & 01:23:22.4 33:25:59 & & & & \phn5010 & \phn15 & \\
 & 01:23:23.5 33:22:55 & & & & \phn4880 & \phn32 & \\
 & 01:23:27.9 33:12:12 & \phn4190 & \phn42 & & \phn4226 & \phn24 & \\
 & 01:23:28.1 33:04:59 & & & & \phn5023 & \phn27 & \\
 & 01:23:28.4 33:19:52 & \phn5922 & \phn20 & & \phn5978 & \phn60 & \\
 & 01:23:37.5 32:37:48 & & & & \phn4748 & \phn20 & \\
 & {\it 01:23:40.0 33:15:20} & & & & \phn4934 & \phn\phn7 & \\
 & 01:23:40.8 33:16:52 & \phn5502 & \phn20 & & \phn5526 & \phn12 & (H$\alpha$, [NII]) \\
 & 01:23:41.6 33:12:10 & & & & \phn5158 & 100 & \\
 & 01:23:43.2 33:24:57 & \phn5650 & \phn46 & & \phn5601 & \phn41 & \\ 
 & 01:23:47.8 33:03:18 & \phn4597 & \phn23 & & \phn4566 & \phn\phn7 & \\ 
 & 01:23:49.5 33:09:22 & \phn4636 & \phn21 & & \phn4537 & \phn60 & \\ 
 & 01:23:50.2 33:35:06 & & & & \phn4984 & \phn26 & \\
 & 01:23:58.5 33:18:46 & \phn5022 & \phn20 & & \phn5056 & \phn27 & \\
 & 01:23:59.1 33:19:51 & \phn4489 & \phn49 & & & & \\
 & 01:23:59.8 33:54:28 & & & & \phn4857 & \phn\phn8 & \\
 & 01:24:01.5 33:51:58 & & & & \phn5145 & \phn\phn9 & \\
 & 01:24:10.9 32:45:58 & & & & \phn6024 & \phn22 & \\
 & 01:24:24.9 33:46:06 & & & & \phn5850 & & \\
 & 01:24:25.7 33:24:24 & \phn4748 & \phn48 & & \phn4983 & \phn39 & \\
 & 01:24:26.8 33:47:58 & & & & \phn5859 & \phn14 & \\
 & 01:24:38.5 33:28:23 & \phn5097 & \phn20 & & \phn5136 & \phn20 & \\ 
 & 01:24:39.5 33:14:09 & & & & \phn5525 & \phn33 & \\
 & 01:24:44.0 33:25:47 & \phn4188 & \phn20 & & \phn4168 & \phn22 & \\
 & 01:25:20.8 34:01:29 & & & & \phn4758 & \phn\phn4 & \\
 & 01:25:33.6 33:40:18 & & & & \phn4879 & \phn28 & \\
 & 01:25:42.7 33:05:16 & & & & \phn4301 & \phn65 & \\
\sidehead{B2 0206+35}
 & 02:08:40.4 35:55:11 & 11129 & \phn20 & & 10818 & \phn62 & \\
 & 02:09:28.8 35:49:25 & 11725 & \phn39 & & & & \\
 & 02:09:29.9 35:38:17 & 11418 & \phn38 & & & & \\
 & 02:09:30.3 36:00:33 & 11231 & \phn74 & & & & \\
 & {\it 02:09:38.7 35:47:49} & 11311 & \phn32 & & 11075 & \phn33 & \\
 & 02:09:44.6 35:54:26 & 10736 & \phn56 & & & & \\
 & 02:09:58.6 35:43:15 & 10752 & \phn37 & & 10790 & \phn26 & (H$\alpha$, [NII]) \\ 
\sidehead{B2 0207+38}
 & {\it 02:10:09.6 39:11:25} & \phn5240 & \phn22 & & \phn5374 & \phn\phn9 & H$\alpha$, [NII], [OII] \\
\sidehead{B2 0222+36}
 & 02:23:55.9 37:03:46 & & & & 10683 & \phn41 & \\
 & {\it 02:25:27.4 37:10:30} & 10009 & \phn20 & & 10009 & \phn30 & (H$\alpha$, [NII], [OI], [OII], [OIII], [SII]) \\
 & 02:25:32.5 36:56:02 & 10951 & \phn31 & & & & \\
 & 02:25:33.0 37:11:25 & 11421 & \phn35 & & & & \\
 & 02:25:34.3 37:03:13 & 10117 & \phn20 & & & & \\
 & 02:25:38.1 36:57:51 & 10753 & \phn29 & & 10807 & \phn29 & \\ 
 & 02:25:45.2 37:13:54 & 11274 & \phn35 & & 11063 & \phn27 & \\ 
 & 02:25:48.1 37:20:52 & 10073 & \phn54 & & & & \\
 & 02:26:12.2 37:13:18 & & & & 11553 & \phn17 & \\ 
 & 02:26:14.3 36:59:13 & 11741 & \phn21 & & & & \\
 & 02:26:14.6 37:17:30 & 10670 & \phn37 & & 10765 & \phn23 & \\ 
 & 02:26:27.6 37:02:25 & 10557 & \phn23 & & 10538 & \phn50 & \\ 
 & 02:26:33.8 37:21:36 & 11332 & \phn51 & & & & \\
 & 02:26:38.7 37:08:10 & 10391 & \phn56 & & & & \\
\sidehead{B2 0258+35}
 & 02:58:03.4 35:15:28 & & & & \phn4850 & \phn\phn5 & \\
 & 03:00:36.4 35:37:45 & & & & \phn4980 & \phn\phn6 & \\
 & 03:00:37.4 35:10:08 & & & & \phn5078 & \phn\phn5 & \\
 & {\it 03:01:42.4 35:12:21} & \phn4926 & \phn20 & & \phn4945 & \phn\phn5 & (H$\alpha$, H$\beta$, [NII], [OI], [OII], [OIII], [SII]) \\
 & 03:01:53.8 35:44:00 & & & & \phn4875 & \phn\phn5 & \\
\sidehead{B2 0326+39}
 & 03:27:33.2 39:53:22 & \phn7634 & \phn35 & & & & \\
 & 03:27:55.2 39:59:51 & \phn7470 & \phn20 & & & & ([NII]) \\
 & 03:28:37.6 39:50:42 & \phn7738 & \phn48 & & & & \\
 & 03:28:53.8 40:02:12 & \phn7225 & \phn22 & & & & ([NII]) \\
 & 03:29:05.0 39:46:00 & \phn7526 & \phn74 & & & & \\
 & 03:29:07.8 39:47:06 & \phn6991 & \phn33 & & \phn6560 & & \\ 
 & {\it 03:29:23.8 39:47:33} & \phn7335 & \phn26 & & \phn7280 & \phn44 & \\ 
 & {\it 03:29:24.7 39:47:54} & & & & \phn7299 & \phn24 & \\
 & 03:29:36.0 39:49:47 & \phn7217 & \phn55 & & & & \\
 & 03:29:51.0 40:01:48 & & & & \phn6805 & & \\
 & 03:30:15.8 39:56:48 & \phn7915 & \phn42 & & & & \\
 & 03:31:08.5 39:37:46 & & & & \phn6586 & \phn28 & \\
 & 03:31:23.7 39:44:30 & & & & \phn5989 & \phn20 & \\
\sidehead{B2 0331+39}
 & 03:31:08.5 39:37:46 & & & & \phn6586 & \phn28 & \\
 & 03:31:23.7 39:44:30 & & & & \phn5989 & \phn20 & \\
 & 03:33:36.8 39:26:04 & \phn5937 & \phn35 & & & & \\
 & {\it 03:34:18.4 39:21:24} & \phn6173 & \phn20 & & \phn6103 & \phn41 & ([NII]) \\ 
 & 03:34:19.3 39:32:44 & \phn6088 & \phn31 & & \phn6058 & \phn38 & \\ 
 & 03:34:19.5 39:35:38 & \phn5538 & \phn39 & & & & \\
 & 03:34:32.5 39:24:01 & \phn6115 & \phn60 & & & & \\
 & 03:34:50.7 39:40:05 & \phn6253 & \phn67 & & & & \\
\sidehead{B2 1317+33}
 & 13:18:59.3 32:58:29 & 11462 & \phn47 & & & & \\
 & 13:20:08.4 33:05:20 & 11813 & \phn46 & & 11754 & \phn48 & \\
 & 13:20:14.5 33:08:38 & 10878 & \phn35 & & 10838 & \phn40 & \\
 & {\it 13:20:17.6 33:08:44} & 11223 & \phn25 & & 11224 & \phn30 & \\
 & 13:20:18.5 33:00:55 & 11112 & \phn58 & & & & \\
 & 13:20:23.7 33:10:39 & 10225 & \phn37 & & & & \\
 & 13:20:31.7 33:17:30 & 10654 & \phn25 & & 10648 & \phn36 & \\
 & 13:20:41.2 33:06:04 & 10852 & \phn28 & & & & \\
 & 13:20:48.8 33:06:04 & 10309 & \phn39 & & & & \\
 & 13:20:53.8 33:08:23 & 10143 & \phn76 & & & & \\ 
 & 13:21:02.4 33:20:25 & 11045 & \phn44 & & 10880 & \phn29 & \\
 & 13:21:31.6 32:55:14 & 11436 & \phn49 & & & & \\
 & 13:21:45.7 33:11:01 & 11179 & \phn39 & & & & \\
 & 13:22:09.7 33:10:08 & 11182 & \phn49 & & & & (H$\alpha$, [NII], [OII], [OIII]) \\
\sidehead{B2 1318+34}
 & {\it 13:20:35.4 34:08:21} & \phn7030 & \phn25 & & \phn6892 & \phn25 & H$\alpha$, H$\beta$, [NII], [OI], [OII], [OIII], [SII] \\
\sidehead{B2 1321+31}
 & 13:20:21.5 31:30:53 & & & & \phn5027 & \phn\phn8 & \\
 & 13:20:30.4 32:00:14 & & & & \phn5521 & \phn22 & \\
 & 13:20:51.8 31:21:59 & & & & \phn5071 & \phn53 & \\
 & 13:21:12.8 31:13:17 & & & & \phn5082 & \phn\phn7 & \\
 & 13:21:19.9 32:08:23 & & & & \phn5456 & & \\
 & 13:21:40.5 31:21:02 & & & & \phn5051 & \phn52 & \\
 & 13:22:51.1 31:49:33 & \phn5289 & \phn33 & & & & H$\alpha$, [NII], [OII], [OIII], [SII] \\
 & 13:23:20.3 32:03:45 & & & & \phn5024 & \phn61 & \\
 & 13:23:41.5 31:38:43 & & & & \phn5016 & \phn42 & \\
 & 13:23:42.9 31:30:55 & \phn5332 & \phn35 & & & & \\
 & {\it 13:23:44.6 31:33:53} & \phn4876 & \phn29 & & \phn4782 & \phn22 & \\
 & 13:24:15.8 31:20:42 & & & & \phn4987 & \phn\phn9 & \\
 & 13:24:33.5 31:40:17 & \phn4697 & \phn35 & & & & \\
 & 13:27:19.4 31:47:18 & & & & \phn4625 & \phn45 & \\
\sidehead{B2 1322+36}
 & 13:24:21.2 35:54:50 & \phn5600 & \phn58 & & \phn5541 & \phn32 & H$\alpha$, H$\beta$, [NII], [OII], [SII] \\
 & 13:24:51.4 36:16:32 & \phn5563 & \phn45 & & \phn5574 & \phn78 & H$\alpha$, [OII], [OIII], [SII] \\
 & {\it 13:24:51.5 36:22:46} & \phn5210 & \phn20 & & \phn5318 & \phn49 & \\
 & 13:25:01.2 36:23:55 & \phn5617 & \phn24 & & \phn5224 & \phn31 & \\
 & 13:25:01.3 36:26:14 & \phn5892 & \phn39 & & \phn5804 & \phn33 & H$\alpha$, H$\beta$, [NII], [OII], [OIII], [SII] \\
 & 13:25:33.4 35:55:27 & \phn5474 & \phn39 & &          &        & H$\alpha$, H$\beta$, [NII], [OII], [OIII], [SII] \\
 & 13:25:36.4 36:22:53 & \phn5639 & \phn58 & & \phn5673 & \phn80 & (H$\alpha$, [NII]) \\
 & 13:26:09.2 35:56:05 & \phn5648 & \phn33 & & \phn5660 & \phn\phn8 & (H$\alpha$--H$\gamma$, [NII], [OII], [OIII], [SII]) \\
 & 13:26:28.6 36:00:37 & \phn5640 & \phn48 & & \phn5533 & 100 & \\
\sidehead{B2 1422+26}
 & 14:24:08.9 26:38:12 & 11544 & \phn30 & & 11418 & \phn45 & ([NII]) \\
 & 14:24:23.8 26:41:25 & 10966 & \phn25 & & 11010 & \phn49 & (H$\alpha$, [NII]) \\
 & {\it 14:24:40.5 26:37:31} & 11138 & \phn20 & & 11090 & \phn26 & (H$\alpha$, [NII], [OI], [OII], [OIII]) \\
 & 14:24:47.3 26:17:19 & 11268 & \phn35 & & & & \\
 & 14:25:02.6 26:41:55 & 11357 & \phn25 & & & & (H$\alpha$, [NII], [OII]) \\
\sidehead{B2 1447+27}
 & 14:48:44.8 27:42:10 & \phn9149 & \phn55 & & & & (H$\alpha$) \\ 
 & 14:49:07.1 27:39:54 & \phn9205 & \phn48 & & & & \\
 & 14:49:08.1 28:02:01 & & & & \phn9620 & \phn50 & \\
 & 14:49:16.1 27:46:36 & & & & \phn9204 & 100 & \\
 & 14:49:25.0 27:46:22 & \phn9332 & \phn22 & & & & \\
 & {\it 14:49:28.0 27:46:50} & \phn9170 & \phn22 & & \phn9231 & \phn24 & \\
 & 14:49:28.1 27:45:21 & \phn9302 & \phn32 & & & & \\
 & 14:49:30.1 27:52:39 & \phn9624 & \phn20 & & \phn9615 & \phn28 & \\
 & 14:49:31.3 27:48:35 & \phn8892 & \phn31 & & & & \\
 & 14:49:54.3 27:42:03 & \phn8914 & \phn20 & & & & H$\alpha$, H$\beta$, [NII], [SII] \\
 & 14:50:02.3 27:49:27 & \phn9325 & \phn29 & & & & \\
 & 14:50:55.4 27:34:43 & \phn9134 & \phn39 & & \phn9053 & \phn33 & \\
\sidehead{B2 1621+38}
 & 16:21:08.8 38:02:00 & 10092 & \phn44 & & & & \\
 & 16:21:30.3 37:51:34 & \phn8629 & \phn64 & & & & \\
 & 16:21:33.9 38:00:31 & 10091 & \phn29 & & 10117 & \phn44 & \\
 & 16:21:42.5 37:45:36 & \phn9871 & \phn20 & & \phn9883 & \phn\phn9 & H$\alpha$--H$\gamma$, [NII], [OII], [OIII], [SII] \\
 & 16:21:43.4 37:59:47 & \phn9962 & \phn25 & & 10022 & \phn43 & \\
 & 16:22:06.5 37:59:59 & \phn9009 & \phn46 & & & & (H$\alpha$, [NII]) \\
 & 16:22:26.1 37:40:54 & \phn9657 & \phn31 & & & & \\
 & 16:22:35.0 38:22:00 & & & & \phn8735 & \phn36 & \\
 & 16:22:43.9 38:09:55 & \phn9227 & \phn36 & & & & \\
 & 16:22:44.2 37:43:08 & \phn9313 & \phn30 & & & & \\
 & 16:22:47.3 37:33:16 & & & & \phn9088 & \phn36 & \\
 & 16:22:48.0 38:02:03 & \phn9136 & \phn31 & & & & (H$\alpha$, H$\beta$, [NII], [SII]) \\
 & 16:22:50.1 37:43:18 & \phn9359 & \phn20 & & & & \\
 & 16:22:50.8 37:46:04 & \phn9181 & \phn65 & & & & (H$\alpha$, [NII]) \\
 & 16:22:53.2 38:14:45 & \phn8471 & \phn20 & & & & \\
 & 16:22:56.4 37:58:58 & \phn9629 & \phn20 & & & & \\
 & 16:22:59.7 37:56:56 & \phn9623 & \phn24 & & \phn9768 & \phn26 & \\
 & 16:22:59.7 38:09:59 & \phn9052 & \phn21 & & & & \\
 & 16:23:00.7 38:05:19 & \phn9242 & \phn55 & & & & \\
 & 16:23:01.3 38:04:41 & \phn8533 & \phn64 & & & & \\
 & {\it 16:23:03.4 37:55:20} & \phn9246 & \phn21 & & \phn9303 & \phn14 & \\
 & 16:23:05.1 37:52:48 & \phn9538 & \phn37 & & & & \\
 & 16:23:14.3 37:55:41 & \phn9873 & \phn29 & & & & \\
 & 16:23:27.9 37:49:52 & \phn8606 & \phn22 & & & & \\
\sidehead{B2 1652+39}
 & 16:52:42.6 39:34:33 & 10119 & \phn30 & & & & H$\alpha$--H$\delta$, [NII], [OII], [OIII], [SII] \\
 & 16:52:52.2 39:51:08 & \phn9694 & \phn31 & & & & \\
 & 16:53:23.1 39:43:01 & 10132 & \phn38 & & & & \\
 & 16:53:48.8 39:46:11 & & & & 10472 & 100 & \\
 & 16:53:51.5 39:48:37 & & & & \phn9622 & 100 & \\
 & {\it 16:53:52.4 39:45:35} & 10036 & \phn27 & & 10092 & \phn22  & \\
 & 16:53:54.0 39:45:31 & & & & \phn9952 & 100 & \\
 & 16:53:56.0 39:44:58 & & & & \phn9982 & 100 & \\
 & 16:53:56.4 39:48:46 & 10018 & \phn20 & & \phn9872 & 100 & \\ 
 & 16:54:01.2 39:44:35 & & & & 10092 & 100 & \\
 & 16:54:06.4 39:46:41 & & & & \phn9992 & 100 & \\
 & 16:54:28.3 39:49:27 & \phn9742 & \phn48 & & & & \\
\sidehead{B2 2116+26}
 & 21:17:42.3 25:57:46 & & & & \phn4569 & \phn\phn4 & \\
 & 21:17:53.4 26:21:17 & \phn4439 & \phn40 & & & & \\
 & {\it 21:18:33.3 26:26:46} & \phn4735 & \phn20 & & \phn4687 & \phn21 & ([NII]) \\
 & 21:18:47.7 26:30:24 & \phn4373 & \phn38 & & & & \\
 & 21:18:54.7 26:29:35 & \phn4741 & \phn51 & & & & \\
\sidehead{B2 2236+35}
 & 22:37:28.1 35:31:34 & \phn8488 & \phn25 & & \phn8633 & \phn29 & (H$\alpha$, [NII], [OII]) \\ 
 & 22:37:40.4 34:50:47 & & & & \phn6940 & \phn29 & \\
 & 22:37:55.0 35:24:14 & \phn7783 & \phn22 & & & & \\
 & 22:38:03.2 35:30:09 & \phn7750 & \phn42 & & & & \\
 & 22:38:12.5 35:13:26 & \phn8204 & \phn80 & & & & (H$\alpha$, [NII]) \\
 & 22:38:13.2 35:29:56 & \phn8070 & \phn20 & & \phn8070 & \phn11 & ([NII]) \\
 & 22:38:25.2 35:21:56 & \phn7857 & \phn20 & & \phn7793 & \phn23 & \\
 & 22:38:26.2 35:22:41 & & & & \phn8634 & \phn28 & \\
 & {\it 22:38:29.4 35:19:50} & \phn8271 & \phn22 & & \phn8318 & \phn27 & \\
 & 22:38:30.2 35:22:42 & \phn8130 & \phn20 & & \phn8104 & \phn31& \\
 & 22:38:34.0 35:23:35 & \phn7624 & \phn35 & & & & \\
 & 22:38:34.8 35:20:30 & \phn8797 & \phn24 & & \phn8789 & \phn27 & \\
 & 22:38:34.8 35:25:40 & \phn8977 & \phn22 & & & & \\
 & 22:38:37.5 35:21:24 & \phn7552 & \phn46 & & & & \\
 & 22:38:44.7 35:32:25 & \phn9352 & \phn20 & & \phn7800 & \phn60 & \\
 & 22:38:58.8 35:26:19 & \phn9124 & \phn22 & & \phn9030 & \phn27 & \\
 & 22:39:25.0 35:42:23 & \phn8173 & \phn56 & & & & (H$\alpha$, H$\beta$, [NII], [OII]) \\
 & 22:39:55.8 35:37:49 & \phn9970 & \phn34 & & & & H$\alpha$, H$\beta$, [NII], [OII], [OIII] \\
\sidehead{3C31}
 & 01:03:26.4 32:14:14 & & & & \phn5319 & \phn\phn9 & \\
 & 01:03:30.5 32:19:04 & & & & \phn5869 & \phn22 & \\
 & 01:04:46.2 32:58:03 & & & & \phn4044 & \phn30 & \\
 & 01:05:19.0 31:40:55 & & & & \phn5865 & \phn10 & \\
 & 01:05:34.2 32:25:47 & & & & \phn5039 & \phn25 & \\
 & 01:05:34.5 31:58:16 & & & & \phn4975 & \phn38 & \\
 & 01:05:50.5 32:23:18 & \phn4592 & \phn28 & & & & \\
 & 01:05:56.0 32:24:43 & \phn4719 & \phn20 & & \phn4706 & \phn\phn6 & \\ 
 & 01:06:45.7 33:13:57 & & & & \phn4473 & \phn60 & \\
 & 01:06:51.4 32:42:05 & & & & \phn5945\tablenotemark{a} & \phn48 & \\
 & 01:06:53.7 32:34:42 & & & & \phn5558\tablenotemark{a} & 116 & \\
 & 01:06:58.2 32:18:30 & \phn5502 & \phn20 & & \phn5581 & \phn25 & \\ 
 & 01:07:03.6 32:23:22 & & & & \phn5096 & \phn\phn7 & \\
 & 01:07:05.7 32:47:42 & & & & \phn5045\tablenotemark{a} & \phn35 & \\
 & 01:07:05.9 32:20:53 & \phn5855 & \phn20 & & \phn6011 & \phn40 & \\ 
 & 01:07:10.3 32:20:48 & \phn4245 & \phn56 & & & & \\
 & 01:07:11.5 32:10:13 & \phn4696 & \phn39 & & \phn4821 & \phn52 & \\ 
 & 01:07:16.0 32:31:12 & \phn5578 & \phn29 & & \phn5492 & \phn31 & \\ 
 & 01:07:17.5 32:28:56 & \phn4421 & \phn20 & & \phn4414 & \phn25 & \\ 
 & 01:07:18.0 32:25:28 & \phn5661 & \phn20 & & & & \\
 & 01:07:23.7 32:24:14 & \phn5229 & \phn20 & & \phn5228 & \phn27 & \\ 
 & 01:07:24.2 32:24:29 & & & & \phn4777 & & \\
 & 01:07:24.9 32:17:31 & \phn4233 & \phn20 & & \phn4287 & \phn23 & \\ 
 & {\it 01:07:25.0 32:24:47} & \phn5008 & \phn20 & & & & \\ 
 & 01:07:27.1 32:19:12 & \phn4975 & \phn20 & & \phn5071 & \phn23 & \\ 
 & 01:07:31.3 32:21:43 & \phn5556 & \phn20 & & \phn5024 & \phn24 & \\ 
 & 01:07:33.0 32:23:27 & & & & \phn4824\tablenotemark{a} & \phn33 & \\
 & 01:07:37.0 32:56:16 & & & & \phn5789 & \phn\phn9 & \\
 & 01:07:46.1 32:41:35 & & & & \phn5434\tablenotemark{a} & \phn31 & \\
 & 01:07:47.1 32:18:37 & \phn5444 & \phn20 & & \phn5534 & \phn25 & \\ 
 & 01:08:02.9 31:40:30 & & & & \phn5551 & \phn33 & \\
 & 01:08:12.9 32:27:13 & \phn4754 & \phn20 & & \phn4846 & \phn26 & \\ 
 & 01:08:15.8 32:29:56 & & & & \phn5661\tablenotemark{a} & \phn75 & \\
 & 01:08:17.2 32:05:24 & & & & \phn4652\tablenotemark{a} & \phn69 & \\
 & 01:08:18.7 32:13:35 & & & & \phn5315\tablenotemark{a} & \phn92 & \\
 & 01:08:23.3 33:08:01 & & & & \phn4697 & \phn27 & \\
 & 01:08:26.0 33:08:53 & & & & \phn4423 & \phn23 & \\
 & 01:08:31.1 33:06:33 & & & & \phn4934 & \phn31 & \\
 & 01:08:43.9 32:06:06 & & & & \phn5718\tablenotemark{a} & \phn87 & \\
 & 01:08:53.6 32:30:52 & \phn4987 & \phn39 & & \phn4978 & \phn28 & \\ 
 & 01:08:55.2 32:19:30 & & & & \phn4735\tablenotemark{a} & \phn59 & \\
 & 01:08:58.9 32:14:25 & & & & \phn4514\tablenotemark{a} & \phn93 & \\
 & 01:08:59.2 32:38:03 & & & & \phn5167 & \phn20 & \\
 & 01:09:13.5 31:58:46 & & & & \phn5255\tablenotemark{a} & \phn25 & \\
 & 01:09:14.0 32:09:04 & & & & \phn4757\tablenotemark{a} & \phn47 & \\
 & 01:09:14.1 32:45:07 & & & & \phn5090 & \phn11 & \\
 & 01:09:33.2 32:10:23 & & & & \phn5606\tablenotemark{a} & \phn31 & \\
 & 01:09:42.4 32:27:04 & & & & \phn5081\tablenotemark{a} & \phn56 & \\
 & 01:09:59.3 32:22:05 & & & & \phn5247\tablenotemark{a} & \phn27 & \\
 & 01:10:20.5 32:23:16 & & & & \phn5395 & \phn27 & \\
 & 01:10:51.8 32:35:35 & & & & \phn5615 & \phn25 & \\
 & 01:11:11.3 32:42:07 & & & & \phn5153 & \phn24 & \\
\sidehead{3C 293}
 & {\it 13:52:17.8 31:26:46} & & & & 13501 & \phn34 & \\
\sidehead{3C 296}
 & 14:15:45.9 10:26:20 & \phn7739 & \phn23 & & \phn7655 & \phn57 & (H$\alpha$, [NII]) \\ 
 & 14:16:08.8 10:35:44 & \phn7931 & \phn29 & & & & \\
 & 14:16:17.3 10:44:56 & \phn6958 & \phn49 & & & & (H$\alpha$) \\
 & 14:16:29.0 10:53:12 & \phn7060 & \phn27 & & & & \\
 & 14:16:30.9 10:37:58 & \phn6666 & \phn21 & & & & \\
 & 14:16:32.7 10:50:50 & \phn7555 & \phn30 & & & & \\
 & 14:16:35.7 10:53:44 & \phn7316 & \phn29 & & & & \\
 & 14:16:42.1 10:42:18 & \phn7346 & \phn65 & & & & \\
 & 14:16:43.2 10:53:10 & \phn7766 & \phn21 & & \phn7740 & \phn45 & \\
 & 14:16:46.2 11:00:29 & \phn8044 & \phn23 & & & & H$\alpha$, H$\beta$, [NII], [OII], [OIII], [SII] \\
 & 14:16:52.6 10:49:36 & \phn7898 & \phn38 & & & & \\
 & {\it 14:16:52.8 10:48:33} & \phn7406 & \phn21 & & \phn7367 & \phn47 & \\
 & 14:16:54.3 10:43:13 & \phn7153 & \phn54 & & & & \\
 & 14:17:01.5 10:47:27 & \phn7006 & \phn20 & & & & \\
 & 14:17:11.9 10:36:43 & \phn6884 & \phn47 & & & & \\
 & 14:17:16.1 10:49:52 & \phn7053 & \phn46 & & & & H$\alpha$, H$\beta$, [NII], [OII] \\
 & 14:17:40.5 10:35:00 & \phn8176 & \phn44 & & & & \\
 & 14:17:41.8 10:45:52 & \phn6913 & \phn49 & & & & \\
 & 14:17:44.1 10:43:55 & \phn7347 & \phn67 & & & & \\
 & 14:18:18.6 11:13:01 & & & & \phn7590 & \phn55 & \\
 & 14:18:22.6 11:11:44 & & & & \phn7399 & \phn51 & \\
\sidehead{3C 305}
 & {\it 14:49:21.5 63:16:15} & 12483 & \phn23 & & 12314 & \phn39 & (H$\alpha$, H$\beta$, [NII], [OI], [OII], [OIII]) \\
 & 14:49:40.5 63:17:55 & 12576 & \phn61 & & & & (H$\alpha$, [NII]) \\
 & 14:50:29.1 63:31:48 & 12567 & \phn34 & & & & \\
\sidehead{3C 386}
 & 18:37:54.5 17:32:02 & & & & \phn4500 & \phn\phn5 & \\
 & {\it 18:38:26.2 17:11:48} & \phn5062 & \phn25 & & \phn5070 & \phn30 & \\
 & 18:38:34.0 17:02:22 & \phn5318 & \phn22 & & & & \\
\sidehead{1651+351}
 & 16:15:41.3 35:08:18 & & & & 10335 & \phn40 & \\
 & 16:16:15.2 34:44:01 & & & & \phn9599 & \phn40 & \\
 & 16:16:28.5 34:39:11 & & & & \phn8688 & \phn40 & \\
 & 16:16:34.4 34:39:46 & & & & \phn9484 & \phn40 & \\
 & 16:16:41.6 35:13:07 & & & & \phn9972 & \phn40 & \\
 & 16:16:45.9 35:17:05 & & & & \phn8947 & \phn40 & \\
 & 16:16:48.6 35:20:23 & & & & \phn8823 & \phn40 & \\
 & 16:16:53.8 34:55:57 & & & & 10458 & \phn40 & \\
 & 16:17:05.3 35:00:10 & \phn9591 & \phn58 & & & & \\
 & 16:17:09.3 34:52:44 & & & & \phn8654 & \phn40 & \\
 & 16:17:12.2 34:52:55 & & & & \phn8697 & \phn40 & \\
 & 16:17:14.6 34:43:20 & & & & 10398 & \phn40 & \\
 & 16:17:20.0 34:54:08 & \phn9399 & \phn20 & & \phn9191 & \phn16 & \\
 & 16:17:24.1 34:58:13 & & & & \phn9163 & \phn40 & \\
 & 16:17:25.7 35:08:08 & \phn9310 & \phn40 & & \phn9116 & \phn19 & \\
 & 16:17:29.6 34:49:22 & \phn9951 & \phn20 & & \phn9918 & \phn40 & \\
 & 16:17:35.6 35:08:03 & \phn9645 & \phn20 & & \phn9712 & \phn31 & \\
 & 16:17:38.7 35:03:36 & & & & \phn9890 & \phn40 & \\
 & 16:17:39.6 35:13:51 & & & & \phn8719 & \phn39 & \\
 & {\it 16:17:40.5 35:00:14} & \phn8979 & \phn20 & & \phn8857 & \phn15 & \\
 & 16:17:43.5 34:57:54 & \phn8098 & \phn52 & & & & \\
 & 16:17:43.9 35:05:13 & \phn9204 & \phn20 & & \phn9218 & \phn40 & \\
 & 16:17:45.1 35:10:45 & & & & \phn9656 & \phn40 & \\
 & 16:17:45.9 34:49:42 & & & & 10172 & \phn40 & \\
 & 16:17:51.2 35:09:24 & \phn9202 & \phn36 & & \phn9214 & \phn40 & H$\alpha$, [NII], [SII] \\
 & 16:17:52.8 35:14:49 & \phn8859 & \phn42 & & \phn8951 & \phn40 & \\
 & 16:18:00.5 35:06:37 & & & & \phn9319 & \phn19  & \\
 & 16:18:01.1 35:15:58 & \phn8892 & \phn21 & & \phn8907 & \phn40 & (H$\alpha$, [NII]) \\
 & 16:18:10.9 34:46:56 & \phn9480 & \phn47 & & & & \\
 & 16:18:16.5 35:10:21 & \phn9871 & \phn22 & & & & \\
 & 16:18:23.7 35:10:26 & & & & \phn8691 & \phn19 & \\
 & 16:18:24.3 35:19:35 & \phn8944 & \phn20 & & \phn8908 & \phn40 & \\
 & 16:18:35.4 35:09:43 & \phn9819 & \phn21 & & & & H$\alpha$, H$\beta$, [NII], [SII] \\
 & 16:18:41.5 34:50:45 & & & & \phn9197 & \phn40 & \\
 & 16:18:54.6 35:09:13 & \phn8854 & \phn20 & & \phn8800 & \phn19 & \\ 
 & 16:18:58.7 34:43:49 & \phn8821 & \phn20 & & \phn8861 & \phn40 & \\
 & 16:19:05.3 35:01:56 & \phn9536 & \phn31 & & \phn9554 & \phn40 & (H$\alpha$, [NII]) \\  
 & 16:19:41.3 35:09:06 & \phn8523 & \phn27 & & \phn8469& \phn40 & \\
\enddata

\tablenotetext{a}{Velocity data from Ledlow et al. (1996).}

\tablecomments{Columns: (1) Field (2) Position, in J2000; the 3C and B2 sources are noted in italics (3) Heliocentric velocity, $cz$, from cross correlation or emission lines (4) Error in velocity (5) Heliocentric velocity, $cz$, from prior studies as reported in NED (6) Error in prior velocity, as reported in NED (7) Emission lines present; if in parentheses the reported velocity is based on cross correlation of the absorption lines.}

\end{deluxetable}

\begin{deluxetable}{c c c c c c c c}
\tablecolumns{8}
\tablecaption{Foreground and Background Galaxy Velocities \label{tbl:bkg_clus}}
\tabletypesize{\small}
\tablewidth{475pt}
\tablehead{
\colhead{} & \colhead{} & \multicolumn{2}{c}{This Study} &
\colhead{} & \multicolumn{2}{c}{Prior Studies} & \colhead{} \\
\cline{3-4} \cline{6-7} \\
\colhead{Field} & \colhead{Position} & \colhead{$cz$} &
\colhead{Error} & \colhead{} & \colhead{$cz$} & \colhead{Error} & 
\colhead{Emission Lines}
}
\startdata
\sidehead{B2 0034+25}
 & 00:36:38.9 25:47:26 & 29152 & \phn48 & & & & \\
 & 00:37:40.1 25:36:47 & \phn7239 & \phn54 & & & & \\
 & 00:37:43.9 25:38:25 & \phn7384 & \phn22 & & \phn7231 & \phn34 & \\
\sidehead{B2 0055+30}
 & 00:55:32.7 30:23:53 & \phn6639 & \phn47 & & \phn6627 & \phn\phn7 & H$\alpha$, H$\beta$, [OII], [OIII] \\
 & 00:55:42.9 30:31:15 & \phn6762 & \phn65 & & & & H$\alpha$, H$\beta$, [NII], [OII], [OIII] \\
 & 00:57:01.9 30:28:59 & \phn6585 & \phn47 & & & & (H$\alpha$, H$\beta$, [NII]) \\
\sidehead{B2 0206+35}
 & 02:09:08.2 35:51:30 & 29066 & \phn47 & & & & \\
 & 02:09:11.0 35:49:43 & 28862 & \phn32 & & & & \\
 & 02:09:50.6 35:28:44 & 13601 & \phn62 & & & & \\
\sidehead{B2 0207+38}
 & 02:09:59.8 39:01:19 & 24147 & \phn38 & & & & \\
 & 02:10:29.7 39:25:02 & 19249 & \phn70 & & & & \\
 & 02:10:33.2 39:08:03 & 25514 & \phn40 & & & & \\
 & 02:10:42.9 38:59:49 & \phn2667 & \phn68 & & & & \\
\sidehead{B2 0222+36}
 & 02:25:17.5 36:48:52 & \phn8385 & \phn72 & & & & \\
\sidehead{B2 0258+35}
 & 03:01:38.2 35:31:45 & 14167 & \phn35 & & & & \\
 & 03:01:39.8 35:16:22 & 12736 & \phn23 & & & & \\
 & 03:01:57.1 35:09:47 & 12737 & \phn27 & & & & \\
 & 03:02:17.1 35:41:02 & 14886 & \phn49 & & & & \\ 
 & 03:02:39.2 35:20:29 & 14315 & \phn44 & & & & \\
\sidehead{B2 0326+39}
 & 03:27:59.3 39:54:17 & \phn4285 & \phn25 & & \phn4337 & \phn14 & H$\alpha$, H$\beta$, [NII], [OIII], [SII] \\
 & 03:28:48.7 40:10:06 & \phn4785 & \phn41 & & & & \\
\sidehead{B2 0331+39}
 & 03:33:57.0 39:04:22 & 16584 & \phn29 & & & & \\
 & 03:34:51.6 39:28:01 & 28764 & \phn52 & & & & \\
 & 03:35:20.1 39:20:19 & 12917 & \phn52 & & & & \\
\sidehead{B2 1318+34}
 & 13:18:50.8 34:01:26 & 17019 & \phn69 & & & & (H$\alpha$, [OII]) \\
 & 13:19:48.1 34:14:11 & 11784 & \phn51 & & & & H$\alpha$, H$\beta$, [OIII] \\
 & 13:19:58.4 34:02:12 & \phn8776 & \phn76 & & & & H$\alpha$, [NII], [OII], [SII]  \\
 & 13:20:09.6 34:02:28 & 24438 & \phn57 & & & & \\
 & 13:20:14.2 33:54:00 & 11635 & \phn25 & & & & H$\alpha$, H$\beta$, [NII], [OI], [OII], [OIII] \\
 & 13:20:31.1 33:58:47 & 11471 & 106 & & & & H$\alpha$--$\gamma$, [NII], [OII], [OIII] \\
 & 13:20:34.0 33:56:43 & 11605 & \phn47 & & & & \\
 & 13:20:37.7 34:11:28 & 19162 & \phn90 & & & & \\
 & 13:20:40.8 34:11:05 & 19417 & \phn72 & & & & H$\beta$, [OII], [OIII] \\
 & 13:20:46.9 34:11:36 & 19513 & \phn75 & & & & ([OII]) \\
 & 13:20:56.0 33:46:06 & 10545 & \phn41 & & & & H$\alpha$, [NII], [OII] \\
 & 13:21:25.0 34:18:35 & 19137 & \phn38 & & & & \\
\sidehead{B2 1321+31}
 & 13:24:18.4 31:55:27 & 23063 & \phn36 & & & & \\
 & 13:24:33.2 31:47:49 & 23246 & \phn33 & & & & \\
 & 13:24:33.7 31:32:46 & \phn6973 & \phn23 & & & & \\
 & 13:25:09.8 32:00:51 & 12237 & \phn20 & & & & H$\alpha$, H$\beta$, [NII], [OII] \\
 & 13:25:35.6 31:55:48 & 11275 & \phn35 & & 11512 & \phn30 & H$\alpha$, H$\beta$, H$\gamma$, [NII], [OII], [SII] \\
 & 13:25:38.3 31:59:38 & \phn7424 & \phn28 & & & & H$\alpha$, H$\beta$, [NII], [SII] \\
 & 13:25:57.2 31:37:06 & \phn7152 & \phn24 & & \phn7280 & \phn30 & \\
\sidehead{B2 1422+26}
 & 14:25:03.6 26:51:26 & 18034 & \phn52 & & & & \\
 & 14:25:18.8 26:37:19 & 23413 & \phn72 & & & & \\
 & 14:25:19.5 26:30:26 & 23860 & \phn34 & & & & \\
 & 14:25:22.8 26:27:56 & \phn4802 & \phn26 & & \phn4744 & \phn32 & \\
\sidehead{B2 1447+27}
 & 14:48:35.4 27:35:01 & 15750 & \phn30 & & & & H$\alpha$, [NII], [OII] \\
 & 14:49:15.9 27:30:35 & 24311 & \phn41 & & & & \\
 & 14:49:28.5 27:37:34 & 18022 & \phn20 & & & & H$\alpha$, H$\beta$, [NII], [OII], [OIII]  \\
 & 14:49:33.0 27:57:28 & 15693 & \phn26 & & & & H$\alpha$, [NII], [OII] \\
 & 14:50:18.8 27:55:07 & 14898 & \phn20 & & & & H$\alpha$, [NII] \\
 & 14:50:27.4 27:35:04 & 24404 & \phn56 & & & & \\
\sidehead{B2 1621+38}
 & 16:22:14.1 37:50:10 & 20663 & \phn43 & & & & check \\
\sidehead{B2 1652+39}
 & 16:51:49.7 39:51:43 & 15721 & \phn20 & & & & H$\alpha$, [NII], [OII] \\
 & 16:52:06.8 40:01:25 & 15682 & \phn65 & & & & (H$\alpha$) \\
 & 16:52:46.2 40:00:37 & 20502 & \phn67 & & & & \\
 & 16:52:53.3 40:09:13 & 44369\tablenotemark{a} & \phn61 & & 44369 & 600 & \\
 & 16:53:05.0 40:07:03 & 44501\tablenotemark{a} & \phn39 & & & & \\
 & 16:53:08.0 39:39:37 & 35400 & \phn53 & & & & \\
 & 16:53:12.3 40:03:05 & 43987\tablenotemark{a} & \phn41 & & & & \\
 & 16:53:21.2 39:57:25 & 20681 & \phn49 & & & & H$\beta$, [OII], [OIII] \\
 & 16:53:40.2 39:47:47 & 35744 & \phn42 & & 35412 & 100 & \\
 & 16:54:00.6 39:53:51 & 11861 & \phn33 & & & & H$\alpha$, H$\beta$, H$\gamma$, [NII], [OII], [OIII], [SII] \\
 & 16:54:04.1 39:57:07 & 44608\tablenotemark{a} & \phn51 & & & & \\
 & 16:54:25.7 39:43:31 & 20787 & \phn68 & & & & \\
 & 16:54:43.3 40:02:47 & 44201\tablenotemark{a} & \phn42 & & & & \\ 
 & 16:54:57.0 39:57:55 & 44531\tablenotemark{a} & \phn60 & & & & \\
 & 16:55:01.4 39:59:28 & 44459\tablenotemark{a} & \phn76 & & & & \\ 
\sidehead{3C 31}
 & 01:07:32.6 32:05:34 & 14569 & \phn76 & & 14851 & \phn35 & \\
\sidehead{3C 293}
 & 13:49:07.8 31:02:02 & 24408 & \phn47 & & & & \\
 & 13:49:08.7 30:51:39 & 24379 & \phn68 & & & & \\
 & 13:49:09.8 30:46:12 & 24333 & \phn42 & & & & \\
 & 13:49:21.9 30:50:13 & 24283 & \phn33 & & & & \\
 & 13:49:24.5 30:59:17 & 24708 & \phn30 & & & & \\
 & 13:49:37.8 30:46:05 & 14961 & \phn25 & & & & \\
 & 13:49:39.4 31:03:26 & 24555 & \phn51 & & & & \\
 & 13:49:46.8 30:53:13 & 24035 & \phn34 & & & & \\
 & 13:49:54.4 30:48:38 & 18807 & \phn20 & & & & H$\alpha$, H$\beta$, H$\gamma$, [OII], [OIII] \\
 & 13:50:37.9 31:20:11 & 13086 & \phn48 & & & & \\
\sidehead{3C 305}
 & 14:47:10.3 63:25:27 & 15481 & \phn33 & & & & \\
 & 14:47:32.6 63:31:09 & 22879 & \phn72 & & & & \\
 & 14:48:06.1 63:13:00 & 15424 & \phn27 & & & & \\
 & 14:48:22.2 63:30:37 & 22798 & \phn36 & & & & \\
 & 14:48:25.4 63:10:10 & \phn8650 & \phn20 & & & & H$\alpha$, H$\beta$, [OII], [OIII] \\
 & 14:48:50.7 63:38:09 & 31921 & \phn46 & & & & (H$\beta$, [OII], [OIII]) \\
 & 14:49:43.0 63:37:36 & 18273 & \phn70 & & & & \\
 & 14:49:50.7 63:44:16 & 27014 & \phn27 & & & & \\
 & 14:50:19.5 63:23:16 & 23280 & \phn47 & & & & \\ 
 & 14:52:02.4 63:38:44 & 14848 & \phn29 & & & & H$\alpha$, H$\beta$, [NII], [OII] \\
 & 14:52:15.9 63:18:35 & 15056 & \phn74 & & & & (H$\alpha$, [OII]) \\
\sidehead{3C 386}
 & 18:38:07.8 17:27:59 & 27593 & \phn60 & & & & \\
\sidehead{1615+351}
 & 16:17:58.6 34:54:37 & 44754 & \phn35 & & & & \\
 & 16:18:24.3 34:58:41 & 23846 & \phn35 & & & & \\
 & 16:18:35.7 34:58:37 & 36435 & \phn55 & & & & (H$\beta$, [OII], [OIII]) \\
\enddata

\tablenotetext{a}{These galaxies are members of an identified cluster from the ROSAT All-Sky Survey, RX J1652.6+4011 \citep{bohr2000}. The cluster is also in the Abell catalog, as Abell 2235.}

\tablecomments{This table contains only velocities measured in the present study. Prior velocity measurements for these galaxies have been taken from NED (when available) and are also included. Columns: (1) Field (2) Position, in J2000 (3) Heliocentric velocity, $cz$, from cross correlation or emission lines (4) Error in velocity (5) Heliocentric velocity, $cz$, from prior studies as reported in NED (6) Error in prior velocity, as reported in NED (7) Emission lines present; if in parentheses the reported velocity is based on cross correlation of the absorption lines.}

\end{deluxetable}

\begin{deluxetable}{l c c c c c c c c c}
\tablecolumns{10}
\tablecaption{Completeness of Velocity Data \label{tbl:complete}}
\tabletypesize{\small}
\tablewidth{0pt}
\tablehead{
\colhead{ } & \colhead{} & \multicolumn{2}{c}{1 Mpc} & \colhead{ } & 
\multicolumn{2}{c}{500 kpc} & \colhead{ } & \multicolumn{2}{c}{250 kpc}\\
\cline{3-4} \cline{6-7} \cline{9-10}
\colhead{Field} & \colhead{Radio} & \colhead{Limit} & 
\colhead{RG+2} & \colhead{} & \colhead{Limit} & \colhead{RG+2} & \colhead{} & 
\colhead{Limit} & \colhead{RG+2}
}
\startdata
B2 0034+25 & 13.7 & 15.1 &   6/7 & & 15.1 &   4/5 & & 15.9 & 1/1 \\
B2 0055+30 & 11.7 & 15.4 &   1/1 & & 15.5 &   1/1 & & 16.0 & 1/1 \\
B2 0120+33 & 11.2 & 14.7 &   3/3 & & 15.4 &   2/2 & & 15.7 & 1/1 \\
B2 0206+35 & 13.3 & 15.6 &   3/3 & & 15.8 &   2/2 & & 15.9 & 2/2 \\
B2 0207+38 & 12.8 & 15.3 &   2/2 & & 16.2 &   2/2 & & 16.2 & 1/1 \\
B2 0222+36 & 14.6 & 15.1 & 15/25 & & 16.6 &   8/9 & & 16.6 & 3/3 \\
B2 0258+35 & 12.4 & 14.7 &   6/6 & & 15.2 &   1/1 & & 15.3 & 1/1 \\
B2 1317+33 & 13.3 & 14.6 &   6/9 & & 14.6 &   3/4 & & 14.6 & 2/3 \\
B2 1318+39 & 13.9 & 14.8 &   3/6 & & 16.0 &   3/3 & & 16.0 & 2/2 \\
B2 1321+31 & 12.8 & 15.0 &   2/2 & & 15.6 &   1/1 & & 16.0 & 1/1 \\
B2 1322+36 & 13.0 & 15.0 &   5/5 & & 15.9 &   3/3 & & 17.2 & 3/3 \\
B2 1422+26 & 14.5 & 15.9 &  5/10 & & 15.9 &   4/7 & & 15.9 & 2/3 \\
B2 1447+27 & 14.5 & 15.6 & 12/16 & & 16.2 & 10/12 & & 16.4 & 6/6 \\
B2 1621+38 & 13.2 & 15.1 &   4/5 & & 16.3 &   3/3 & & 16.9 & 2/2 \\
B2 1652+39 & 13.8 & 15.1 &   3/5 & & 16.2 &   3/3 & & 16.2 & 2/2 \\
B2 2236+35 & 13.1 & 15.2 &   8/8 & & 15.2 &   7/7 & & 15.2 & 5/5 \\
3C 31      & 12.3 & 14.9 &   5/5 & & 15.6 &   3/3 & & 16.3 & 3/3 \\
3C 293     & 14.6 & 16.7 &   1/1 & & 17.0 &   1/1 & & 17.3 & 1/1 \\
3C 296     & 12.5 & 15.4 &   3/3 & & 15.4 &   2/2 & & 15.5 & 2/2 \\
3C 305     & 14.5 & 15.9 &   1/1 & & 17.0 &   1/1 & & 17.5 & 1/1 \\
1615+351   & 14.0 & 15.8 & 25/27 & & 15.8 & 15/17 & & 15.8 & 5/6 \\
\enddata

\tablecomments{Columns: (1) Radio galaxy field. (2) $R_c$ magnitude of the radio galaxy, taken from the APS. The radio galaxy represents the brightest galaxy in the field for all sources except B2 0222+36, B2 1322+36, and 1615+351. (3) $R_c$ magnitude below which all galaxies within 1 Mpc projected separation from the radio galaxy have measured velocities. (4) Fraction of galaxies within 2 magnitudes of the radio galaxy which have measured velocities. (5--6) Same as columns 3 and 4 but for galaxies with 500 kpc of the radio galaxy. (7--8) Same as columns 3 and 4 but for galaxies within 250 kpc of the radio galaxy.}

\end{deluxetable}

\begin{deluxetable}{l r c c c c c r c c c}
\tablecaption{Summary Table \label{tbl:sumtab}}
\tabletypesize{\tiny}
\tablewidth{405pt}
\tablehead{
\colhead{ } & \multicolumn{5}{c}{1 Mpc} & \colhead{ } & \multicolumn{4}{c}{250 kpc} \\
\cline{2-6} \cline{8-11}
\colhead{Field} & \colhead{N} & \colhead{cz} & \colhead{$\sigma$} & \colhead{$\Delta_{RG}$} & \colhead{M} &
\colhead{ } & \colhead{N} & \colhead{cz} & \colhead{$\sigma$} & \colhead{$\Delta_{RG}$} \\
\colhead{ } & \colhead{ } & \colhead{[km s$^{-1}$]} & \colhead{[km s$^{-1}$]} & \colhead{ } & \colhead{[$\times 10^{14}$ M$_\odot$]} &
\colhead{ } & \colhead{} & \colhead{[km s$^{-1}$]} & \colhead{[km s$^{-1}$]} & \colhead{ } 
}
\startdata
B2 0034+25 & 11 & \phn9875$\pm$182 & 507$^{+\phn97}_{-\phn82}$ & $+1.8$ & 1.37$^{+0.53}_{-0.44}$ &  & 4  & \phn9595$\pm$179 & 468$^{+688}_{-277}$ & $+0.3$ \\
B2 0055+30 & 13 & \phn4903$\pm$138 & 267$^{+103}_{-\phn74}$ & $-0.5$ & 0.83$^{+0.64}_{-0.46}$ & & 4  & \phn5186$\pm$343 & 308$^{+999}_{-285}$ & $+0.6$ \\
B2 0120+33 & 65 & \phn4984$\pm$\phn67 & 591$^{+\phn44}_{-\phn41}$ & $+0.7$ & 3.07$^{+0.46}_{-0.43}$ &  & 19 & \phn4967$\pm$115 & 504$^{+\phn88}_{-\phn75}$ & $+0.3$ \\
B2 0206+35 & 7 & 11203$\pm$150 & 360$^{+106}_{-\phn82}$ & $-0.7$ & 1.11$^{+0.66}_{-0.51}$ & & 3 & 11272$\pm$372 & 415$^{+388}_{-201}$ & $-0.1$ \\
B2 0207+38 & 1 & \nodata & \nodata & \nodata & \nodata & & 1 & \nodata & \nodata & \nodata \\
B2 0222+36 & 14 & 10810$\pm$161 & 575$^{+\phn92}_{-\phn79}$ & $+4.9$ & 2.85$^{+0.91}_{-0.79}$ & & 3 & 11345$\pm$591 & 154$^{+918}_{-132}$ & $+2.3$ \\
B2 0258+35 & 5 & \phn4932$\pm$\phn60 & \phn86$^{+\phn53}_{-\phn33}$ & $+0.1$ & 0.08$^{+0.11}_{-0.06}$ & & 2 & \nodata & \nodata & \nodata \\
B2 0326+39 & 13 & \phn7290$\pm$137 & 477$^{+175}_{-128}$ & $-0.1$\tablenotemark{a} & 1.14$^{+0.83}_{-0.61}$ & & 5 & \phn7295$\pm$\phn57 & 188$^{+131}_{-\phn77}$ & $-0.1$\tablenotemark{a} \\
B2 0331+39 & 8 & \phn6102$\pm$\phn63 & 266$^{+211}_{-118}$ & $-1.1$ & 0.47$^{+0.74}_{-0.41}$ & & 3 & \phn6098$\pm$\phn94 & 113$^{+237}_{-\phn77}$ & $-0.8$ \\
B2 1317+33 & 14 & 11017$\pm$143 & 504$^{+129}_{-103}$ & $-1.4$ & 1.99$^{+1.02}_{-0.81}$ & & 5 & 10961$\pm$194 & 554$^{+331}_{-207}$ & $-1.3$ \\
B2 1318+34 & 1  & \nodata & \nodata & \nodata & \nodata & & 1 & \nodata & \nodata & \nodata \\
B2 1321+31 & 14 & \phn5051$\pm$\phn66 & 268$^{+244}_{-128}$ & $+2.4$ & 0.74$^{+1.35}_{-0.71}$ & & 4 & \phn4961$\pm$152 & 247$^{+213}_{-114}$ & $+0.5$ \\
B2 1322+36 & 6\tablenotemark{b} & \phn5625$\pm$\phn27 & \phn30$^{+\phn29}_{-\phn15}$ & \nodata & 0.01$^{+0.01}_{-0.01}$ & & 3 & \phn5611$\pm$\phn30 & \phn35$^{+\phn58}_{-\phn22}$ & \nodata \\
B2 1422+26 & 5 & 11258$\pm$112 & 209$^{+105}_{-\phn70}$ & $+1.1$ & 0.38$^{+0.38}_{-0.25}$ & & 2 & \nodata & \nodata & \nodata \\
B2 1447+27 & 12 & \phn9213$\pm$\phn45 & 225$^{+\phn88}_{-\phn63}$ & $+0.9$ & 0.24$^{+0.19}_{-0.14}$ & & 7 & \phn9197$\pm$\phn90 & 251$^{+\phn93}_{-\phn68}$ & $+0.3$ \\
B2 1621+38 & 24 & \phn9290$\pm$\phn93 & 502$^{+\phn70}_{-\phn62}$ & $+0.5$ & 2.07$^{+0.58}_{-0.51}$ & & 7 & \phn9438$\pm$243 & 408$^{+209}_{-138}$ & $+0.8$ \\
B2 1652+39 & 12 & \phn9990$\pm$\phn68 & 216$^{+125}_{-\phn79}$ & $-0.6$ & 0.14$^{+0.17}_{-0.11}$ & & 8 & 10010$\pm$\phn23 & \phn53$^{+151}_{-\phn39}$ & $-0.7$ \\
B2 2116+26 & 5 & \phn4572$\pm$120 & 155$^{+\phn35}_{-\phn29}$ & $-1.3$ & 0.07$^{+0.03}_{-0.03}$ & & 4 & \phn4575$\pm$313 & 179$^{+999}_{-174}$ & $-0.5$ \\
B2 2236+35 & 18 & \phn8244$\pm$155 & 722$^{+177}_{-142}$ & $-0.2$ & 2.28$^{+1.12}_{-0.90}$ & & 10 & \phn8172$\pm$174 & 486$^{+\phn99}_{-\phn82}$ & $-0.6$ \\
3C 31      & 52 & \phn5120$\pm$\phn63 & 493$^{+\phn39}_{-\phn36}$ & $+1.7$ & 1.76$^{+0.28}_{-0.26}$ & & 18 & \phn5155$\pm$172 & 514$^{+\phn85}_{-\phn73}$ & $+0.8$ \\
3C 293     & 1  & \nodata & \nodata & \nodata & \nodata & & 1 & \nodata & \nodata & \nodata \\
3C 296     & 21 & \phn7377$\pm$\phn98 & 428$^{+\phn59}_{-\phn52}$ & $-0.3$ & 1.17$^{+0.32}_{-0.28}$ & & 10 & \phn7340$\pm$\phn94 & 301$^{+\phn82}_{-\phn64}$ & $-0.7$ \\
3C 305\tablenotemark{c} & 1  & \nodata & \nodata & \nodata & \nodata & & 1 & \nodata & \nodata & \nodata \\
3C 386     & 3  & \phn4995$\pm$319 & 372$^{+546}_{-221}$ & $-0.2$ & 0.97$^{+2.85}_{-1.15}$ & & 2 & \nodata & \nodata & \nodata \\
1615+351   & 38 & \phn9286$\pm$162 & 584$^{+\phn76}_{-\phn67}$ & $+1.9$ & 3.23$^{+0.84}_{-0.74}$ & & 8 & \phn9324$\pm$101 & 417$^{+336}_{-186}$ & $+3.4$ \\
\enddata

\tablenotetext{a}{The radio galaxy is a binary. The other half of the binary is $-0.3\sigma$ (1 Mpc) and $-0.6\sigma$ (250 kpc) different from the group systemic velocity.}

\tablenotetext{b}{The radio galaxy lies just outside this group. An additional two galaxies are likely associated with the system bringing its total to nine, all of which are included in Table \ref{tbl:memvel}. For further discussion, see the text.}

\tablenotetext{c}{3C 305 lies to the edge of a galaxy pair. The radio galaxy and this nearby pair are included in Table \ref{tbl:memvel}.}

\tablecomments{Columns: (1) Field; (2) Number of galaxies with measured velocities consistent with group membership; (3) Systemic velocity, $cz$, of the group, as determined using the biweight estimator of location ($C_{BI}$); (4) Velocity dispersion of the group, as determined using the biweight estimator of scale ($S_{BI}$); (5) Significance of difference in velocity of radio galaxy from systemic velocity of group; (6) Revised virial mass of system; (7) -- (10) same as (2) -- (5) but for galaxies within 250 kpc of the radio galaxy. Note that the results within 250 kpc have not had additional 3$\sigma$ clipping performed. We have not included system masses for the 250 kpc limit, as the low number of velocities makes the uncertainties large enough that they are of little value.}

\end{deluxetable}

\clearpage

\begin{figure}
\figurenum{1}
\epsscale{0.85}
\plotone{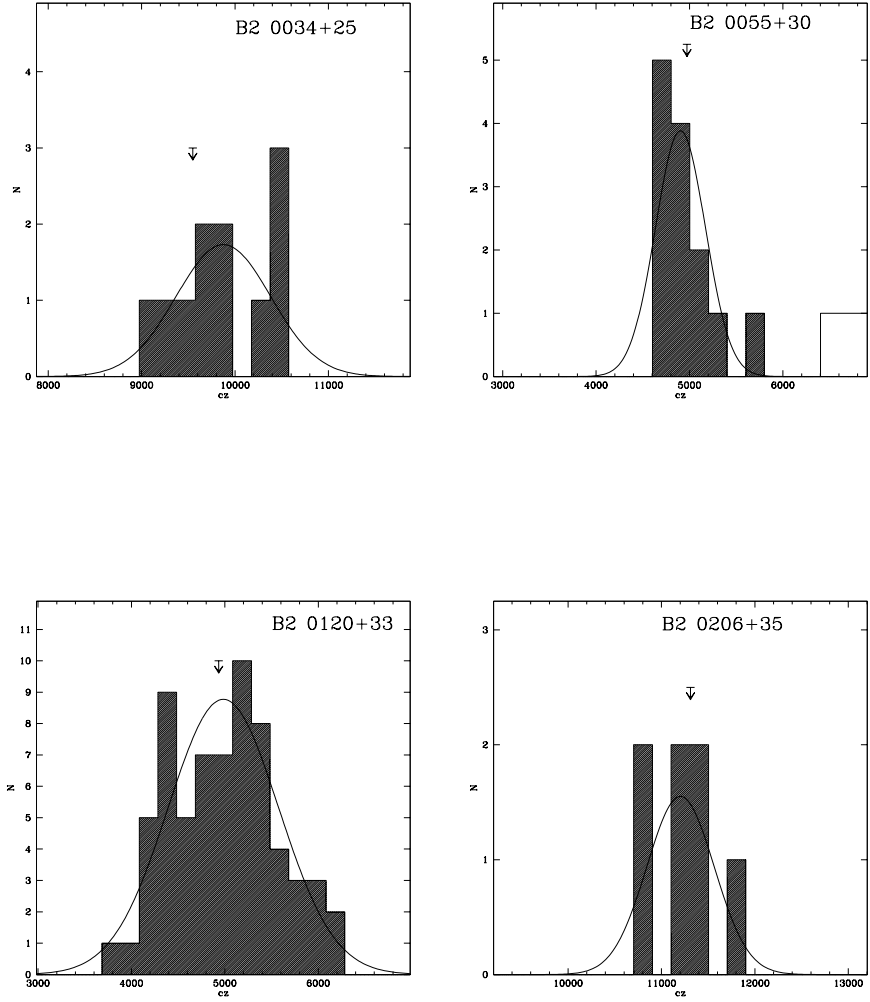}
\caption{Velocity histograms for the 25 sources investigated. For systems consisting of more than one galaxy, a Gaussian with mean and standard deviation corresponding to the derived systemic velocity ($C_{BI}$) and dispersion ($S_{BI}$) is also plotted. Unshaded portions of the histograms represent foreground and background galaxies. The velocity of the radio galaxy is marked with an arrow. \label{fig-1}}
\end{figure}

\begin{figure}
\figurenum{1}
\epsscale{0.85}
\plotone{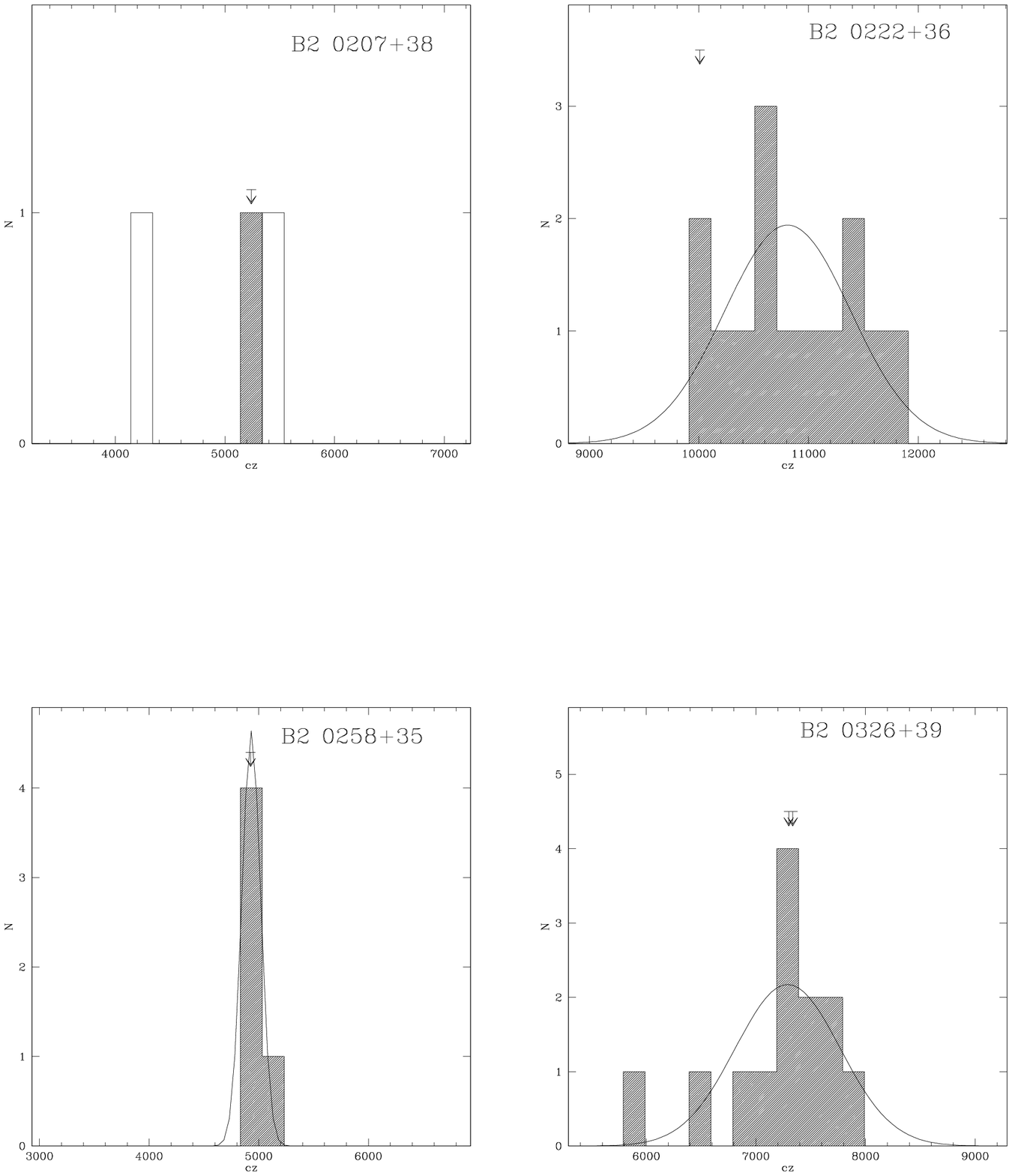}
\caption{Continued.}
\end{figure}

\begin{figure}
\figurenum{1}
\epsscale{0.85}
\plotone{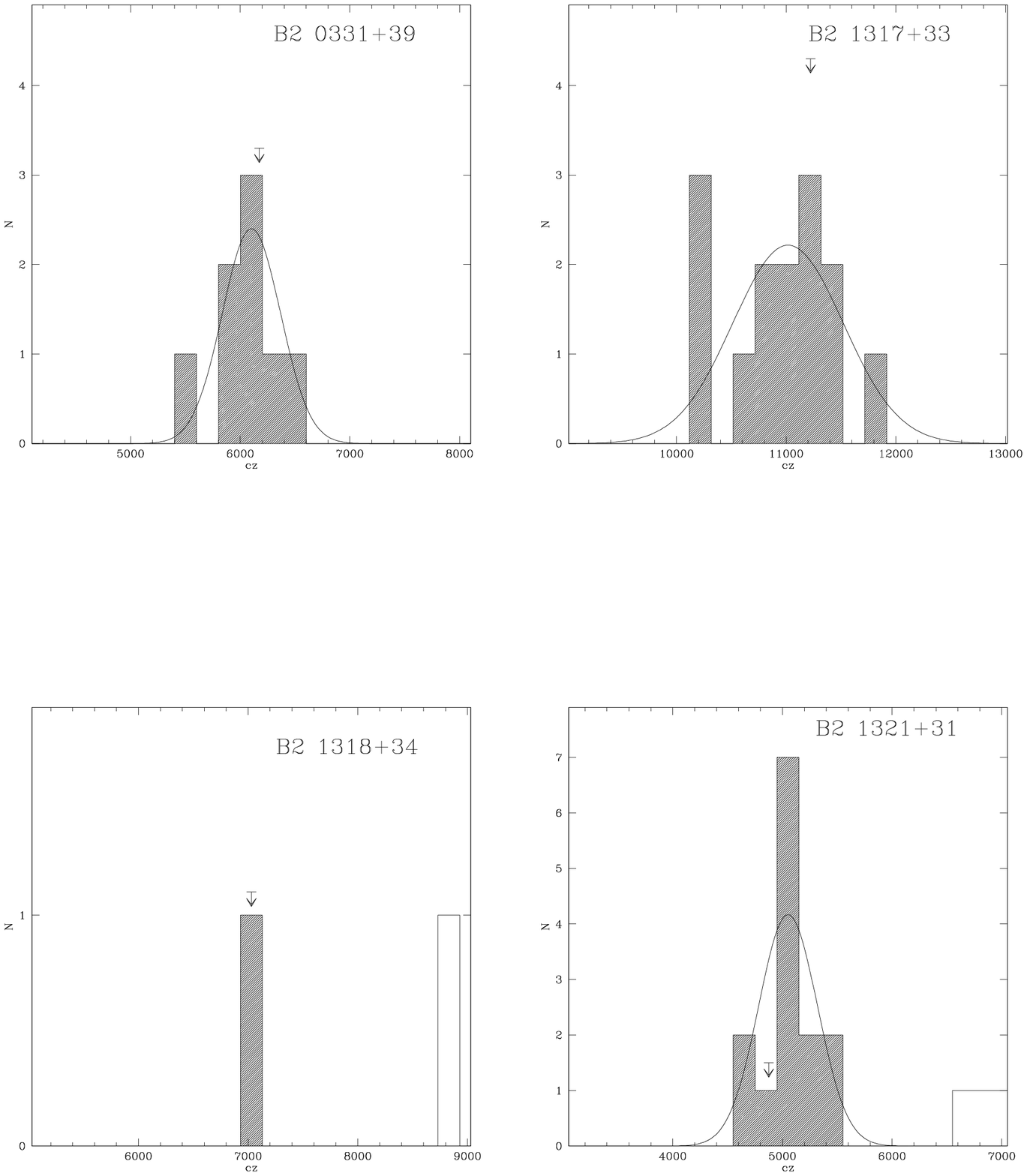}
\caption{Continued.}
\end{figure}

\begin{figure}
\figurenum{1}
\epsscale{0.85}
\plotone{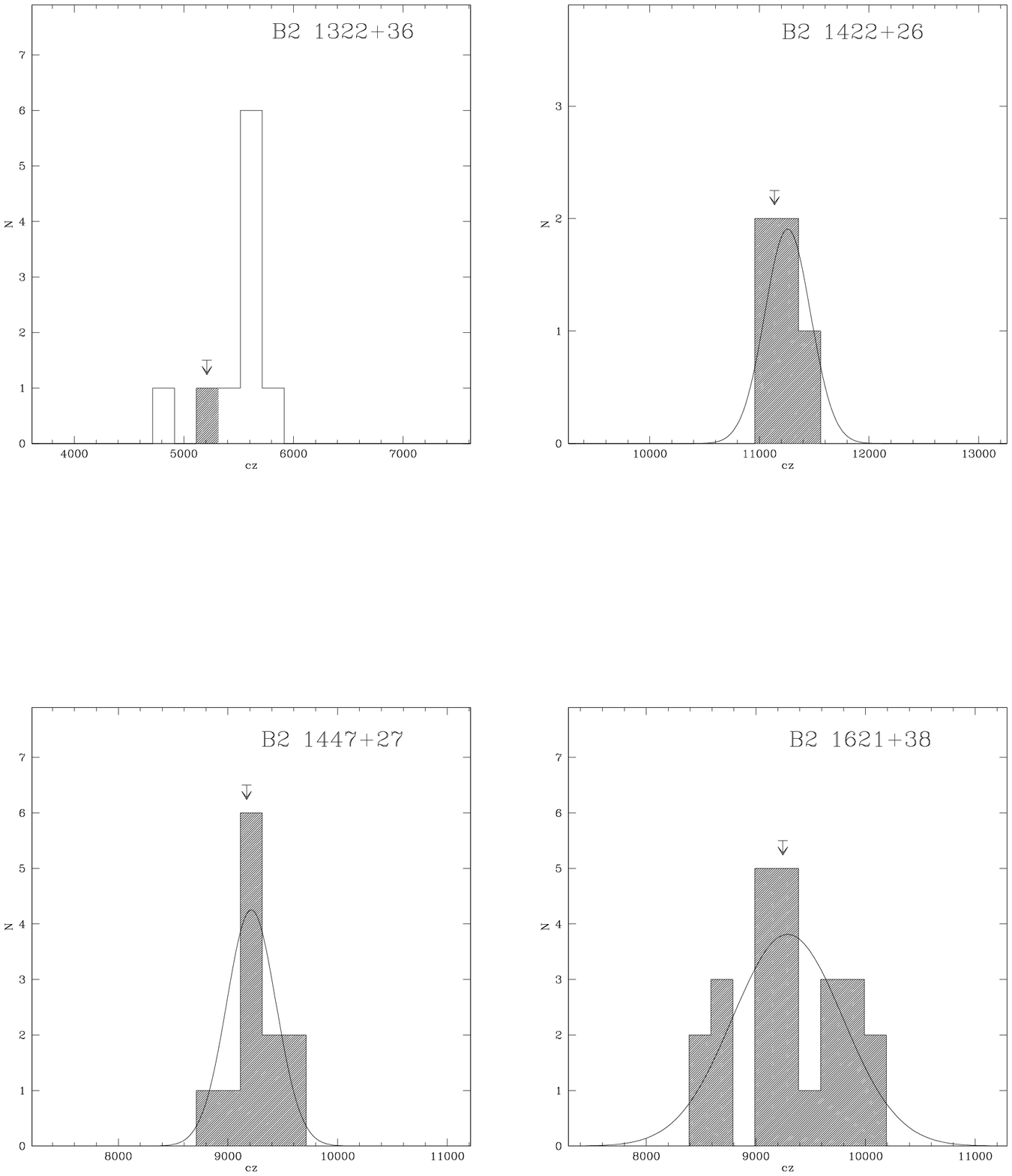}
\caption{Continued.}
\end{figure}

\begin{figure}
\figurenum{1}
\epsscale{0.85}
\plotone{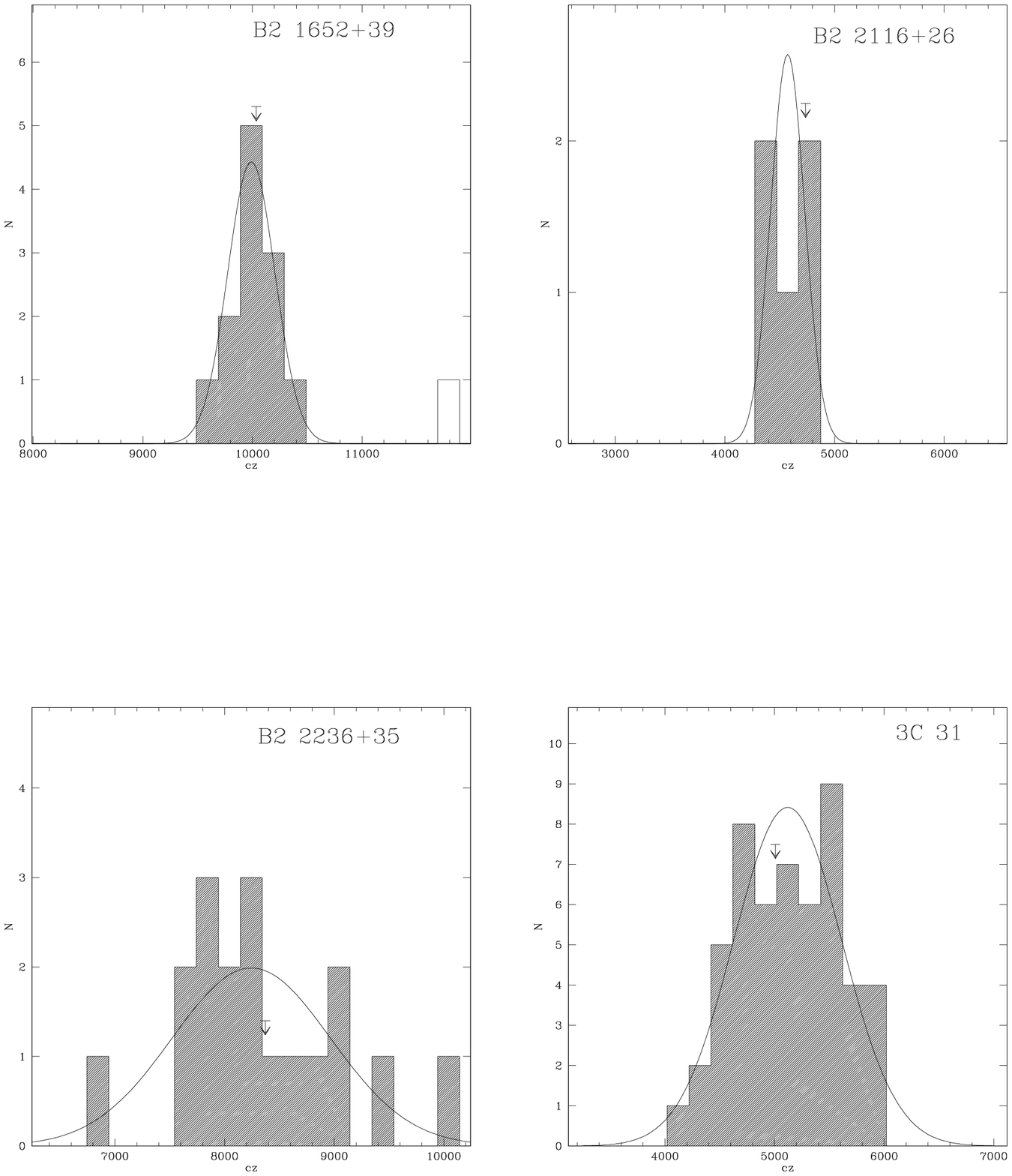}
\caption{Continued.}
\end{figure}

\begin{figure}
\figurenum{1}
\epsscale{0.85}
\plotone{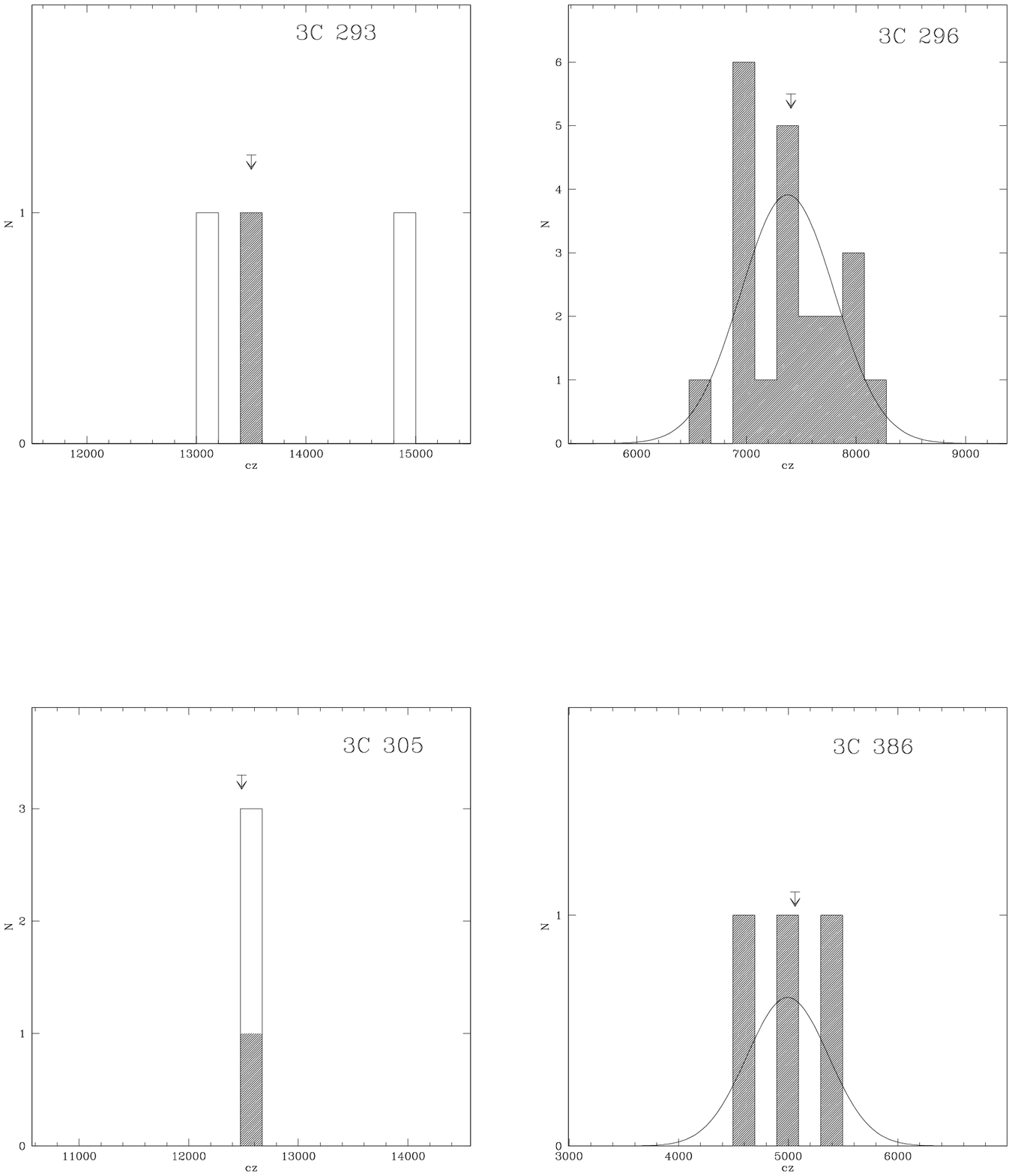}
\caption{Continued.}
\end{figure}

\begin{figure}
\figurenum{1}
\epsscale{0.85}
\plotone{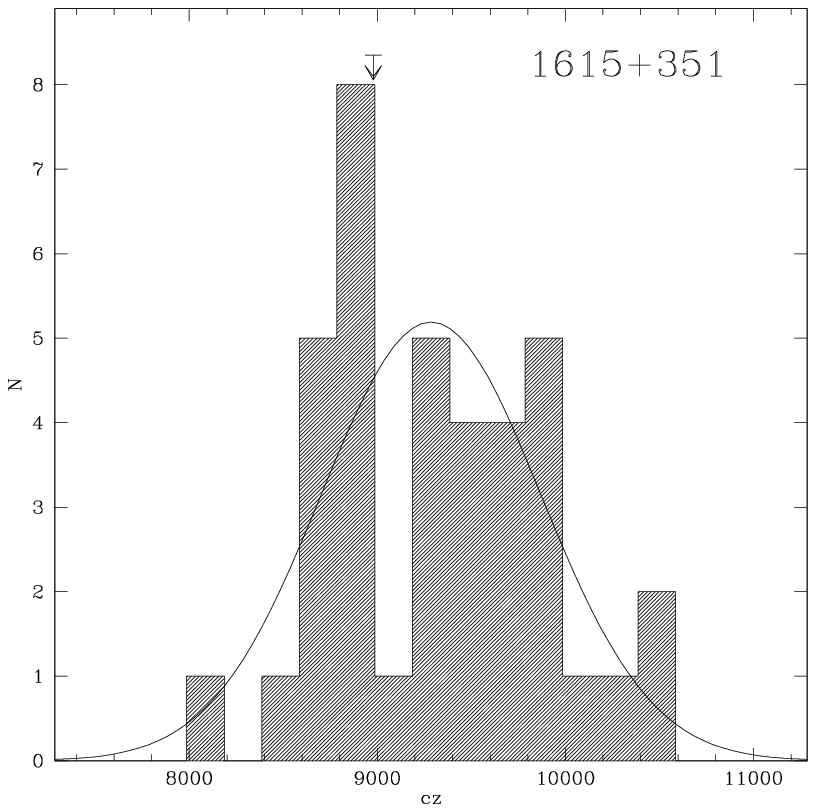}
\caption{Continued.}
\end{figure}

\end{document}